# Tracking X-Ray Variability in Next-generation EHT Low-luminosity Active Galactic Nucleus Targets

Nicole M. Ford[1,2,12], Michael Nowak[3], Venkatessh Ramakrishnan[4,5], Daryl Haggard[1,2], Kristen Dage[6], Dhanya G. Nair[7,8], and Chi-kwan Chan[9,10,11]
[1] Department of Physics, McGill University, 3600 Rue University, Montreal, Québec H3A 2T8, Canada
[2] Trottier Space Institute, 3550 Rue University, Montréal, Québec H3A 2A7, Canada
[3] Department of Physics, Washington University in Saint Louis, MSC 1105-109-02, One Brookings Drive, St. Louis, MO 63130-4899, USA
[4] Finnish Centre for Astronomy with ESO, University of Turku, FI-20014 Turku, Finland
[5] Aalto University Metsähovi Radio Observatory, Metsähovintie 114, FI-02540 Kylmälä, Finland
[6] International Centre for Radio Astronomy Research—Curtin University, GPO Box U1987, Perth, WA 6845, Australia
[7] Astronomy Department, Universidad de Concepción, Casilla 160-C, Concepción, Chile
[8] Max Planck Institute for Radio Astronomy, Auf dem Hügel 69, 53121 Bonn, Germany
[9] Steward Observatory and Department of Astronomy, University of Arizona, 933 N. Cherry Avenue, Tucson, AZ 85721, USA
[10] Data Science Institute, University of Arizona, 1230 N. Cherry Avenue, Tucson, AZ 85721, USA
[11] Program in Applied Mathematics, University of Arizona, 617 N. Santa Rita, Tucson, AZ 85721, USA

## Abstract

We present a 5 month NICER X-ray monitoring campaign for two low-luminosity active galactic nuclei (LLAGNs)—NGC 4594 and IC 1459—with complementary Swift and NuSTAR observations. Utilizing an absorbed power-law and thermal source model combined with NICER's SCORPEON background model, we demonstrate the effectiveness of joint source–background modeling for constraining emission from faint, background-dominated targets. Both sources are dominated by nuclear power-law emission with photon indices $\Gamma \sim 1.5$–2, with NGC 4594 being slightly harder than IC 1459. The thermal contribution in both sources is fainter, but constant, with $kT \sim 0.5$ keV ($\sim 5 \times 10^6$ K). The power-law flux and $\Gamma$ are strongly anticorrelated in both sources, as has been seen for other LLAGNs with radiatively inefficient accretion flows. NGC 4594 is the brighter source and exhibits significant aperiodic variability. Its variability timescale with an upper limit of 5–7 days indicates emission originating from $\lesssim 100\ r_g$, at the scale of the inner accretion flow. A spectral break found at $\sim 6$ keV, while tentative, could arise from synchrotron/inverse Compton emission. This high-cadence LLAGN X-ray monitoring campaign underlines the importance of multiwavelength variability studies for a sample of LLAGNs to truly understand their accretion and outflow physics.

*Unified Astronomy Thesaurus concepts:* Active galaxies (17); X-ray active galactic nuclei (2035); LINER galaxies (925)

## 1. Introduction

In active galactic nuclei (AGNs), the accretion and feedback from the central supermassive black hole (SMBH) directly impact the host galaxy. The brightest AGNs, known as quasars/QSOs and blazars, have been extensively surveyed (see, e.g., M. Schmidt 1963; E. Y. Khachikian & D. W. Weedman 1974; G. T. Richards et al. 2006). These systems are thought to have a thin disk configuration (N. I. Shakura & R. A. Sunyaev 1973). Standard thin disks are optically thick and radiatively efficient, so heat can be radiated away and generate a bright nuclear region. There is evidence, however, that SMBH activity undergoes cyclical "flickering," oscillating between lower and higher accretion/emission states (T. Di Matteo et al. 2005; J. E. Greene & L. C. Ho 2007; K. Schawinski et al. 2015). AGNs with low accretion rates and thus luminosities are hereafter referred to as low-luminosity AGNs (LLAGNs). LLAGNs have typical bolometric luminosities ($L_{\rm bol}$) below $10^{42}$ erg s$^{-1}$ ($10^9\ L_\odot$), as a result of accreting at <0.01 times the Eddington luminosity ($L_{\rm Edd}$). These may actually represent the most common state of AGN activity; some form of LLAGN is thought to exist in aproximately one-third of all galaxies in the local Universe ($z \approx 0$; L. C. Ho 2008). Understanding LLAGN accretion and feedback is therefore essential to interpreting the evolution of both SMBHs and their host galaxies.

In LLAGNs, the inner accretion disk is likely a radiatively inefficient accretion flow (RIAF; R. Narayan 2002). An RIAF is optically thin, geometrically thick, and contains a quasi-spherical inflow in the innermost tens of gravitational radii ($r_g = 2GM/c^2$). The astrophysical community is actively pursuing emission signatures that distinguish LLAGNs as being in an RIAF state (e.g., F. Yuan et al. 2003; R. S. Nemmen et al. 2006; L. C. Ho 2009; P. N. Best & T. M. Heckman 2012; R. She et al. 2018; A. Jana et al. 2023). Meanwhile, a correlation has been observed between low accretion rate and jet-dominated emission (see N. M. Nagar et al. 2005; L. C. Ho 2008 and references therein). In RIAF systems, a substantial fraction of the gravitational energy is released kinetically through jets and outflows, for example via the Blandford–Znajek mechanism (R. D. Blandford & R. L. Znajek 1977). The relative contribution to the emission in the inner hundreds of $r_g$ from an RIAF versus a jet is a matter of active debate (e.g., R. Narayan 1996; S. Markoff et al. 2007; Event Horizon Telescope Collaboration et al. 2019; J. A. Fernández-Ontiveros

---

[12] Corresponding author.

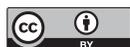

et al. 2023; I. Ruffa et al. 2024), with observers seeking ways of disentangling the two.

The broadband spectral energy distributions (SEDs) of LLAGNs have several distinct features, presumably related to their RIAF and/or jet emission. They tend to be bright in centimeter to radio frequencies (see summary in L. C. Ho 2008). They show a steep power-law continuum extending from the infrared (IR)–optical to the ultraviolet (UV; e.g., S. P. Willner et al. 1985; J. P. Halpern & M. Eracleous 1994; J. A. Fernández-Ontiveros et al. 2023). LLAGNs do not have the expected optical–UV "blue bump," which is the characteristic spectral signature of thermal emission from the accretion disk (M. Chiaberge et al. 2006; L. C. Ho 2008; G. Younes et al. 2012).

In the X-ray regime, LLAGNs can show a thermal bump below 2 keV. This bump may arise from some combination of star formation processes (e.g., O. González-Martín et al. 2009; S. Falocco et al. 2020; Y. Diaz et al. 2023), a hot interstellar medium (e.g., D.-W. Kim et al. 2019; G. Younes et al. 2019; Y. Diaz et al. 2023), and/or warm Comptonization of a truncated standard accretion disk located at tens to hundreds of $r_g$ (e.g., A. Kubota & C. Done 2018; G. Younes et al. 2019; S. Falocco et al. 2020). Emission above 2 keV follows a power-law shape (e.g., L. C. Ho et al. 2001; D. Donato et al. 2004; L. C. Ho 2008; O. González-Martín et al. 2006, 2009, 2016; R. She et al. 2018; J. A. Fernández-Ontiveros et al. 2023). The shape originates from accelerated leptons within the jets and/or the inner accretion flow that Comptonize either locally produced photons (synchrotron self-Compton (SSC)) or externally produced photons (for further details see, e.g., C. D. Dermer et al. 1992; M. Sikora et al. 1994; P. F. Hopkins et al. 2009; M. A. Prieto et al. 2016; B. Bandyopadhyay et al. 2019). LLAGNs lack the relativistic iron reflection lines associated with X-ray emission from, for example, typical Seyfert I disks (see, e.g., A. Ptak et al. 2004; T. Kawamuro et al. 2013). The absence of both relativistic X-ray reflection and optical/UV thermal disk emission has been taken as evidence for LLAGNs being in an RIAF state, potentially with a truncated standard disk (see, e.g., the RIAF models of M81* in A. J. Young et al. 2007, 2018). More multiwavelength observations of a sample of LLAGNs are necessary to fill in the details of this departure from the standard AGN disk structure.

In the past two years, the Event Horizon Telescope (EHT) collaboration has begun submillimeter observations of a selection of LLAGNs. A handful of the highest priority LLAGN targets are identified as part of the Event Horizon and Environs sample (V. Ramakrishnan et al. 2021, 2023; D. G. Nair et al 2024). These select targets have SMBH masses and spins measurable by EHT, and they represent the "next generation" of most likely candidates to display event horizons resolvable by a future expanded EHT array (M. D. Johnson et al. 2023; X. A. Zhang et al. 2024). To inform EHT studies and to further explore the accretion and feedback mechanisms of LLAGNs, we conducted an extensive X-ray monitoring campaign of two high priority EHT + multiwavelength targets: NGC 4594 and IC 1459. Key properties of these sources are reported in Table 1 and summarized below.

NGC 4594 (referred to as Sombrero for the remainder of this work) is a nearby ($z \sim 0.003$) nearly edge-on ($<84°$) spiral galaxy with a central low-ionization nuclear emission-line region (LINER). It has a two-sided jet/counterjet structure extending from a radio core (J. F. Gallimore et al. 2006; K. Hada et al. 2013; M. Mezcua & M. A. Prieto 2014; P. Kharb et al. 2016; Y. Yang et al. 2024) that is inclined at $\sim66°$ (X. Yan et al. 2024). Its 230 GHz flux provided in the Atacama Large Millimeter/submillimeter Array (ALMA) Source Catalog[13] is 198 mJy (0.198 Jy). The central SMBH has a measured mass in the range of $6.6 \times 10^8 M_\odot$ (J. R. Jardel et al. 2011; K. Gültekin et al. 2019) to $1.0 \times 10^9 M_\odot$ (J. Kormendy 1988; J. Kormendy et al. 1996). Its event horizon has a predicted angular size of $\theta_{\rm ring} \sim 7$–10 $\mu$as depending on which mass estimate is used; this is large enough to be potentially resolved by EHT in the future. The broadband SED shows evidence for a mid-infrared "bump" attributed to thermal emission from a cold and/or truncated accretion disk (J. A. Fernández-Ontiveros et al. 2023). Recent measurements of broad mid-infrared emission-line widths with JWST indicate some form of outflow is present, but it is unclear whether this comes from an RIAF-induced wind or jets (there may also be a nuclear dust contribution; see K. Goold et al. 2024 for further details). The nucleus appears as a moderately bright X-ray source with some extended emission attributed to hot gas and/or stellar binaries (Z. Li et al. 2010, 2011). Power-law fits to Chandra (e.g., S. Pellegrini et al. 2003; O. González-Martín et al. 2009; Z. Li et al. 2011), XMM-Newton (S. Pellegrini et al. 2003), BeppoSAX (S. Pellegrini et al. 2002), and ASCA (K. L. Nicholson et al. 1998) spectra from the nucleus generally follow a slope $\Gamma \sim 2$ (Z. Li et al. 2011).

IC 1459 is a well-studied elliptical galaxy with symmetric jets (S. J. Tingay & P. G. Edwards 2015) located at $z \sim 0.006$. It has a counterrotating stellar core (M. Franx & G. D. Illingworth 1988), which may suggest it underwent a merger sometime in its past. The central SMBH has a measured mass in the range of 2.5–2.6 $\times 10^8 M_\odot$ (M. Cappellari et al. 2002; A. Schulze & K. Gebhardt 2011; R. P. Saglia et al. 2016; K. Gültekin et al. 2019). Its event horizon has a predicted angular size of $\theta_{\rm ring} \sim 8.9$–9.0 $\mu$as depending on which mass estimate is used. Its nucleus contains a strong (1 Jy) compact radio source with a diameter less than $0.''03$ (O. B. Slee et al. 1994) and a hard AGN-like X-ray component (ASCA; H. Matsumoto et al. 1997). The 230 GHz flux is 217 mJy (0.217 Jy; I. Ruffa et al. 2019) at arcsec-scale resolution. The intrinsic nuclear X-ray source has a power-law form with fitted slope $\Gamma \sim 1.9$ combined with a thermal hot gas component with $kT \sim 0.6$ keV (G. Fabbiano et al. 2003; L. R. Ivey et al. 2024).

The remainder of this manuscript is organized as follows. The X-ray data and analysis strategy to produce spectra and light curves are presented in Section 2. Spectral fitting and light-curve variability findings, as well as comparisons to existing literature, are reported in Section 3. In Section 4, we interpret the spectral and variability properties in the context of LLAGN accretion and feedback mechanisms. A final summary and conclusions are presented in Section 5.

## 2. Data Reduction and Analysis

In 2022, we monitored three LLAGN targets of interest to EHT using the Neutron star Interior Composition ExploreR (NICER; Program Number 5676; K. C. Gendreau et al. 2012) telescope at 0.2–12 keV energies.[14] This work will focus on the emission characteristics of the two fainter targets—Sombrero

---
[13] https://almascience.nrao.edu/sc/
[14] NICER spectra in the 12–15 keV band are almost exclusively used for background modeling.

Table 1
Selection of EHT High Priority Low-luminosity Active Galactic Nucleus Targets Shown in Bold, Sorted in Descending Order of Average 0.3–20 keV (NICER + NuSTAR) X-Ray Brightness

| Source | Distance (Mpc) | Galactic $N_H$ $10^{20}$ atoms cm$^{-2}$ | log$M_{BH}$ ($M_\odot$) | Ring Size ($\mu$as) | Millimeter Flux Density (Jy) | Average X-Ray Flux (erg cm$^{-2}$ s$^{-1}$) | Inferred $f_{Edd}$ ($10^{-5}$) |
|---|---|---|---|---|---|---|---|
| M87* | 16.8 | 2.54 | 9.8 | 42 | 1.2 | 7.3E–12 | 0.83 |
| **Sombrero** | 9.87 | 3.67 | 8.82 | 7.0 | 0.2 | 1.3E–12 | 1.8 |
| **IC 1459** | 28.92 | 1.15 | 9.4 | 8.9 | 0.2 | 1.0E–12 | 2.4 |
| Sgr A* | 0.008 | 110.10 | 6.6 | 51.8 | 2.4 | 1.8E–13 | 0.056 |

**Note.** M87* and Sgr A*—the two LLAGNs with event horizons already imaged by EHT—are included for comparison purposes, with source properties taken from Event Horizon Telescope Collaboration et al. (2019), GRAVITY Collaboration et al. (2019), and The Event Horizon Telescope Collaboration (2022), and references therein. Average X-ray fluxes are taken from the NASA/IPAC Extragalactic Database. $N_H$ values are taken from the Chandra Galactic Neutral Hydrogen Density Calculator, which queries the Colden database compiled from J. M. Dickey & F. J. Lockman (1990). Black hole masses for Sombrero and IC 1459 are from J. R. Jardel et al. (2011) and M. Cappellari et al. (2002), and distances are from J. Kormendy & L. C. Ho (2013). Inferred $f_{Edd}$ are from X. A. Zhang et al. (2024), see their Section 4.1 for more detail on their calculations.

and IC 1459—with the third target (NGC 3998) presented in a future study (V. Ramakrishnan et al. 2025, in preparation). Sombrero was observed from 2022 March to August and IC 1459 was observed from 2022 May to October. All sources were observed for 3 ks every ∼3 days for each LLAGN during their respective observing windows. A handful of observations for each source had too little exposure and/or too much background noise to be scientifically useful, and are excluded from further analysis. This leaves 49 and 41 usable NICER observations for Sombrero and IC 1459, respectively.

During the NICER X-ray monitoring, we also obtained a snapshot observation of each target with the Nuclear Spectroscopic Telescope Array (NuSTAR; Program Number 00008229; F. A. Harrison et al. 2013) and the Neil Gehrels Swift Observatory (Swift; N. Gehrels et al. 2004). The Swift observations were performed with all three instruments on board: the X-ray Telescope (XRT; D. N. Burrows et al. 2005), the Ultraviolet/Optical Telescope (P. W. A. Roming et al. 2005; 170–600 nm), and the Burst Alert Telescope (S. D. Barthelmy et al. 2005; 15–150 keV). In this work, we focus only on the XRT data. A short EHT + ALMA observation of Sombrero was also taken during the NICER monitoring, and will be the subject of a future study.

All extracted spectra are modeled in the X-ray spectral fitting package XSPEC (K. A. Arnaud 1996).[15] Given the faintness of our sources (∼0.1 counts s$^{-1}$ in the NICER band), we opt to use Cash statistics (W. Cash 1979) when fitting rather than the classic $\chi^2$. Due to low count rates and increased residuals beyond a certain energy threshold, source fluxes are only extracted from fits in the range of 0.3–8 keV for NICER, 3–20 keV for NuSTAR, and 0.3–5 keV for Swift.[16]

### 2.1. NICER

The NICER observations are processed using HEASoft v6.32.1,[17] starting with the NICERL2[18] task. Due to low signal (≲1 counts s$^{-1}$), high overshoots (>30), and signatures of background flaring activity present in a significant fraction of the observations, the overshoots ("overonly" range) are restricted to <1.5 for Sombrero and <1.0 for IC 1459. The undershoots ("underonly" range) are kept at the default level of <500. Overshoots track the non-X-ray high-energy particle backgrounds (e.g., cosmic rays), while undershoots track detector resets due to excess charge from, e.g., optical photons and thermal excitations (B. LaMarr et al. 2016).

#### 2.1.1. Spectra

Spectra are extracted using the NICERL3-spec task, configured with the optimal binning scheme introduced in J. S. Kaastra & J. A. M. Bleeker (2016) and a minimum of 10 counts bin$^{-1}$. The background spectra are generated using the SCORPEON tool[19] with the "script" setting, which enables adjustable background components that can be jointly modeled with the source. The 0.3–15 keV energy range is used to facilitate fitting both the background and source models.

The spectra are fit using XSPEC version 12.13.1, with solar abundances from J. Wilms et al. (2000). The photoelectric absorption cross sections were taken from D. A. Verner et al. (1996). We fix the target galaxies' redshifts to values taken from the NASA/IPAC Extragalactic Database (NASA/IPAC Extragalactic Database (NED) ),[20] and the Galactic absorption $N_H$ to the values reported in the Chandra Galactic Neutral Hydrogen Density Calculator (J. M. Dickey & F. J. Lockman 1990).[21] We extract the 2–10 keV absorbed fluxes for the combined model components using the XSPECcflux convolution model. Due to increased residuals beyond ∼8 keV, individual model components' absorbed fluxes are only extracted in the 0.3–8 keV range.

Previous efforts to fit the soft (here defined as 0.3–8 keV) X-ray spectra of these sources found that both power-law and thermal emission components were needed to describe the spectral shape (e.g., G. Fabbiano et al. 2003; S. Pellegrini et al. 2003; O. González-Martín et al. 2009; Z. Li et al. 2011). The power-law emission may be attributed to inverse Compton scattering of optical–UV disk photons passing through a coronal region.[22] The thermal emission may be attributed to

---

[15] https://heasarc.gsfc.nasa.gov/xanadu/xspec/
[16] IC 1459 was too faint to constrain the spectrum beyond the NICER band.
[17] https://heasarc.gsfc.nasa.gov/docs/software/heasoft/
[18] https://heasarc.gsfc.nasa.gov/docs/nicer/nicer_analysis.html
[19] https://heasarc.gsfc.nasa.gov/docs/nicer/analysis_threads/scorpeon-overview/
[20] NED (https://ned.ipac.caltech.edu/) is funded by the National Aeronautics and Space Administration and operated by the California Institute of Technology.
[21] https://cxc.harvard.edu/toolkit/colden.jsp
[22] In this context, we interpret corona to mean an X-ray-emitting source near the central SMBH, presumed to be made up of a plasma of hot electrons (e.g., F. Haardt & L. Maraschi 1991).

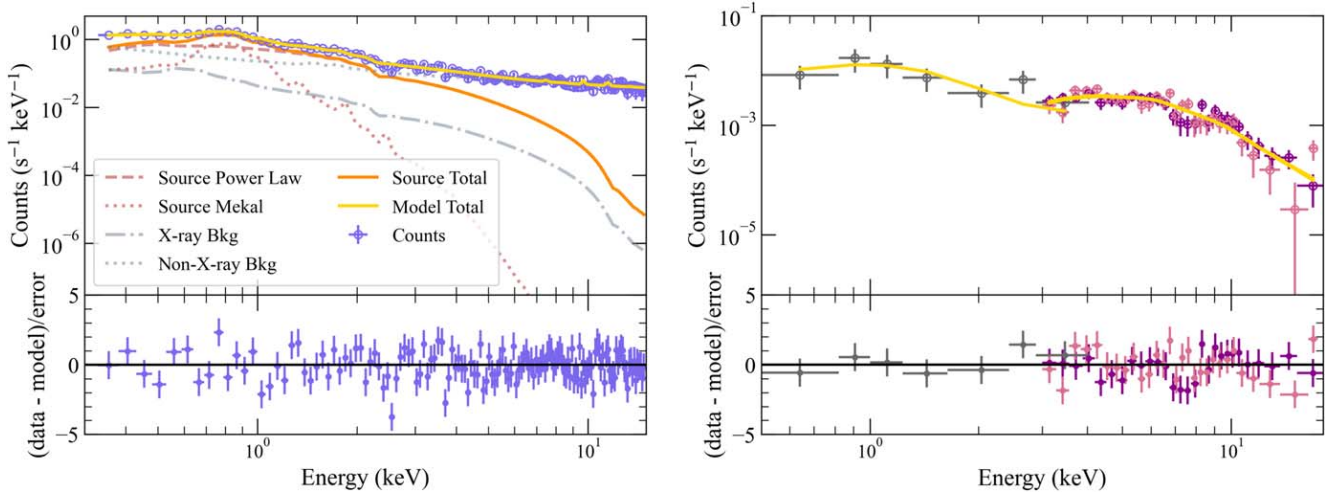

**Figure 1.** Left panel: Sombrero NICER 0.3–15 keV spectrum showing counts (blue), source `powerlaw` component (red dashed line), source `mekal` component (red dotted line), total source model (Equation (1); orange line), X-ray background (gray dashed–dotted line), non-X-ray background (gray dotted line), and combined source + background models (yellow solid line), generated using `SCORPEON` and fit in `XSPEC`. Residuals are shown in the bottom panel. Right panel: Sombrero joint fit of the Swift + NuSTAR 0.5–18 keV background-subtracted source spectrum. NuSTAR Focal Plane Modules A and B (FPMA and FPMB, respectively) are shown with purple and pink points, respectively, and Swift data are shown in dark gray. The joint model (Equation (2), yellow line) and residuals are shown in the bottom panel.

Table 2
Source Model Parameters from Fits to the NICER and, in the Case of Sombrero, Joint Swift + NuSTAR Data

| | Sombrero | | IC 1459 |
|---|---|---|---|
| | NICER | Swift + NuSTAR | NICER[a] |
| $\langle \Gamma_1 \rangle$ | $1.7 \pm 0.4$ | $1.6^{+0.3}_{-0.4}$ | $2.1 \pm 0.7$ |
| $E_{\text{break}}$ [keV] | ⋯ | $6.2^{+1.0}_{-1.7}$ | ⋯ |
| $\Gamma_2$ | ⋯ | $2.9 \pm 0.4$ | ⋯ |
| $\langle kT \rangle$ [keV] | $0.5 \pm 0.1$ | $0.5$[b] | $0.52 \pm 0.09$ |
| $\langle N_H \rangle$ [$10^{22}$ atoms cm$^{-2}$] | $0.02$[c] | $0.02$[c] | $0.1$[c] |
| $\langle F_{\text{pl}} \rangle$ [$10^{-12}$ erg cm$^{-2}$ s$^{-1}$] | $2.5 \pm 0.9$ | $1.14 \pm 0.05$ | $0.9 \pm 0.4$ |
| $\langle F_{\text{th}} \rangle$ [$10^{-12}$ erg cm$^{-2}$ s$^{-1}$] | $0.25 \pm 0.06$ | $<0.2$[c] | $0.24 \pm 0.04$ |
| $\sigma^2_{\text{NEV}}$ | $0.12 \pm 0.05$ | ⋯ | $-0.02 \pm 0.05$ |

**Notes.** Reported values for NICER are the unweighted averages (denoted by brackets) with 1σ standard deviation from all observations for all free parameters, and for Swift + NuSTAR it is the fitted value and the 90% confidence region for the single observation. We also report the average power-law $\langle F_{\text{pl}} \rangle$ and thermal $\langle F_{\text{th}} \rangle$ soft X-ray fluxes (0.3–8 keV) and the normalized excess variance ($\sigma^2_{\text{NEV}}$) for the power-law emission. The thermal emission does not vary enough to produce a meaningful $\sigma^2_{\text{NEV}}$.
[a] The corresponding Swift + NuSTAR snapshot observations have insufficient signal for modeling.
[b] The Swift + NuSTAR fit was insensitive to the thermal component, so we freeze $kT$ to the median value from the NICER observations and only estimate an upper limit on $F_{\text{th}}$.
[c] For both LLAGN targets, the majority of the NICER fits are insensitive to the intrinsic $N_H$; in these cases, we freeze $N_H$ to the median value from all fits, and then refit the observations. For Swift + NuSTAR we freeze $N_H$ to the same value as for NICER.

another warm corona of diffuse gas (though, see, e.g., A. Kubota & C. Done 2018; A. Zoghbi & J. M. Miller 2023, and references therein for alternative explanations). This thermal component produces a bump that dominates the emission at ≲2 keV. While more complex thermal (or disk reflection) models could be invoked, we opted for a simple `mekal` component in `XSPEC` in order to be consistent with several existing studies of these sources. The full model is:

$$\text{tbabs} \times (\text{ztbabs} \times \text{powerlaw} + \text{mekal}). \quad (1)$$

A representative joint source–background fit is shown in the left panel of Figure 1. The `tbabs` component accounts for absorption from the interstellar medium within our Galaxy and should affect all host galaxy emission components. The `ztbabs` component represents the intrinsic local absorption of the nuclear emission by the host galaxy. Both the redshift and the Galactic hydrogen column density are fixed to published values for each source (see Table 1). The intrinsic host galaxy absorption is frequently only constrained with an upper limit in our fits—in these situations, we freeze it to the median value from the other observations (reported in Table 2). We find that $N_H$ is always $\lesssim 0.1 \times 10^{22}$ atoms cm$^{-2}$ for Sombrero, and $\lesssim 2.0 \times 10^{22}$ atoms cm$^{-2}$ for IC 1459 in our spectral fits.

SCORPEON includes customizable models for both the X-ray and non-X-ray backgrounds. For the majority of our joint source–background fits, we keep all the default SCORPEON parameter settings. In cases where the reduced fit statistic was $\gtrsim 1.3$ or $\lesssim 0.9$ and/or the source fit parameters deviate significantly from the general trends, we refit while enabling variations in the constant non-X-ray background ("CON"), the cosmic-ray background ("COR"), and the lower energy solar wind charge exchange O K and O VII lines.

We estimate the uncertainties on all free parameters and allow refits if new minima are found. If a refit is required, we recalculate all the uncertainties following the refit. The steppar tool is used when nonmonotonicity was detected. If any of the SCORPEON terms have unconstrained uncertainties in the fits, we freeze them back to their defaults and refit. We report all fit results with uncertainties for both Sombrero and IC 1459 in Tables 4 and 5.

### 2.1.2. Intraday Light Curves

Because HEASoft v6.32.1 does not allow SCORPEON to be used for background estimation when generating intraday light curves, we also use the updated v6.33.1 provided on the SciServer platform[23] solely to create light curves with the NICERL3-lc task including the SCORPEON background. We first rerun the NICERL3-spec task to generate the appropriate ancillary response files and response matrix files used by NICERL3-lc with SCORPEON. Light curves are extracted in the 0.3–8 keV range with 35 s bins. There is no evidence for significant variability of the source in the intraday observations, with most fluctuations attributable to variations in the background and accounted for by SCORPEON.

### 2.2. Swift

Swift (0.2–10 keV, 18″ resolution, ~24′ field of view) observed both targets quasi simultaneously with NuSTAR during the NICER monitoring campaign. A single ~1 ks XRT observation was performed in photon counting mode for each target. Spectra were generated using the Swift-XRT data product generator tool provided by the UK Swift Science Data Centre (for details, see P. A. Evans et al. 2009), with the iterative centroid option enabled. We applied ftgrouppha in the 0.2–10 keV range using the same optimal binning scheme used on the NICER data, with the exception that we we imposed a condition on the required counts per bin to five instead of 10.

### 2.3. NuSTAR

A single 20 ks NuSTAR (3–79 keV, 18″ resolution, ~10′ field of view) observation was obtained during the NICER monitoring campaign for each target. Level 2 and 3 data products were extracted using nupipeline and nuproducts, respectively, for FPMA and FPMB. A 50″ region was used to select the source, and a nearby 110″ region was used to select the background. We applied ftgrouppha in the 3–79 keV range with the same optimal binning scheme used for the NICER data (see Section 2.1.1).

### 2.4. Joint Swift–NuSTAR Fit

The Swift and NuSTAR observations occurred on the same day for Sombrero, so we perform a joint fit. The quasi-simultaneous NICER observation on that day unfortunately had no good time interval (periods when observing conditions were good), so it was excluded from the joint fit. The observations for IC 1459 have insufficient signal for modeling, so we focus the remainder of this section on Sombrero.

For the joint fit, we changed the Equation (1) model's power-law component to a broken power law, and incorporated a cross-normalization term for each instrument (including NuSTAR FPMA and FPMB). The full model is:

$$\text{const} \times \text{tbabs} \times (\text{ztbabs} \times \text{bknpower} + \text{mekal}). \quad (2)$$

The joint Swift–NuSTAR fit is shown in the right panel of Figure 1, and the fit parameters are reported in the second column of Table 2. Using the binning described in the previous sections, we are able to fit the 0.3–20 keV range. Due to low counts in the Swift spectrum, the thermal gas component is difficult to constrain. We therefore opt to freeze its model parameters to the mean values derived from the NICER data (see Table 2); the thermal emission appears roughly constant in time, so variability in the parameters is not a major concern.

For the joint fit, using a simple power law in place of the broken power law yielded a marginally satisfactory fit (C-statistic/degrees of freedom (DOF) $\sim 76/59$). Including the break leads to an improvement in the fit statistic ($\Delta C \sim 20$) with the addition of only two free parameters. We additionally perform an F-test to gain a rough estimate of the significance of the postbreak $\Gamma_2$, finding a chance probability of $\lesssim 10^{-4}$. The prebreak $\Gamma_1$ also agrees nicely with the $\langle \Gamma \rangle$ found in the NICER fits (see Table 2), whereas the simple power-law $\Gamma$ skews high ($\Gamma = 2.3 \pm 0.1$). We therefore proceed with the broken power-law component in our joint model; the break is interpreted further in Section 4.2.

## 3. Results

We summarize the results of our spectral fits to all observations in Table 2. Example fits to NICER, NuSTAR, and Swift observations of Sombrero are shown in Figure 1. The fitted power-law ($F_{pl}$) and thermal ($F_{th}$) fluxes from NICER are combined to make ~5 month light curves for Sombrero and IC 1459. These light curves are shown in Figure 2 and plotted with comparisons to relevant literature fluxes, along with the corresponding power-law $\Gamma$ and thermal $kT$ from our fits. In this section, we compare the fitted spectra parameters (Table 2) of our two sources with each other and with existing literature, and analyze their variability.

### 3.1. Spectral Modeling: Sombrero

We compare our findings to several existing X-ray studies employing similar power-law and thermal emission models. S. Pellegrini et al. (2003) analyze XMM-Newton (40 ks) and Chandra observations (18.7 ks) and find $F_{pl} \sim 2 \pm 0.1 \times 10^{-12}$ erg cm$^{-2}$ s$^{-1}$ (0.5–10 keV) and $\Gamma \approx 1.9$. O. González-Martín et al. (2006, 2009) reanalyze the Chandra and XMM-Newton data. Their best Chandra fit is a basic power law with $F_{pl}$ comparable to S. Pellegrini et al. ([2003](#)). They report $\Gamma \approx 1.4$ (O. González-Martín et al. [2006](#)) and $\Gamma \approx 1.6$ (O. González-Martín et al. [2009](#)). For the XMM-Newton data they fit a combined power-law and thermal model, finding $\Gamma \approx 2$. Z. Li et al. ([2011](#)) fit all existing Chandra observations of Sombrero taken between 1999 and 2008 (including those used in S. Pellegrini et al. [2003](#); O. González-Martín et al. [2006](#), [2009](#)), finding $\langle F_{pl} \rangle \sim 2 \times 10^{-12}$ erg cm$^{-2}$ s$^{-1}$ and $\Gamma \sim 1.9 \pm 0.1$. These fluxes are on average comparable to those found in our fits, while $\Gamma$ is slightly steeper but within one standard deviation of our average. For the thermal component, S. Pellegrini et al. ([2003](#)) find $F_{th} \sim 2.6 \pm 0.9 \times 10^{-14}$ erg cm$^{-2}$ s$^{-1}$ (0.5–2 keV) and $kT \sim 0.6$ keV. In the XMM-Newton fit by O. González-Martín et al. ([2009](#)), they find comparable fluxes and temperatures. Z. Li et al. ([2011](#)) find $F_{th}$ is constant at $\sim 0.2 \times 10^{-12}$ erg cm$^{-2}$ s$^{-1}$ (0.3–2 keV) with $kT \sim 0.58$ keV. Our results agree closely with O. González-Martín et al. ([2006](#), [2009](#)) and Z. Li et al. ([2011](#)), and generally agree with S. Pellegrini et al. ([2003](#)), though their $F_{th}$ is slightly lower. Based on the analyses in Z. Li et al. ([2010](#)) and Z. Li et al. ([2011](#)), the combined flux from nonnuclear discrete X-ray sources should be subdominant to $F_{pl}$; $F_{th}$ likely includes a contribution from the discrete sources, but our NICER modeling is not sensitive enough to distinguish those components. Additionally, the intrinsic $N_H$ in all these studies is $\sim 10^{21}$ atoms cm$^{-2}$, which is in agreement with the upper limit of our NICER fits, indicating that Sombrero is only moderately absorbed.

---

[23] https://apps.sciserver.org/dashboard/

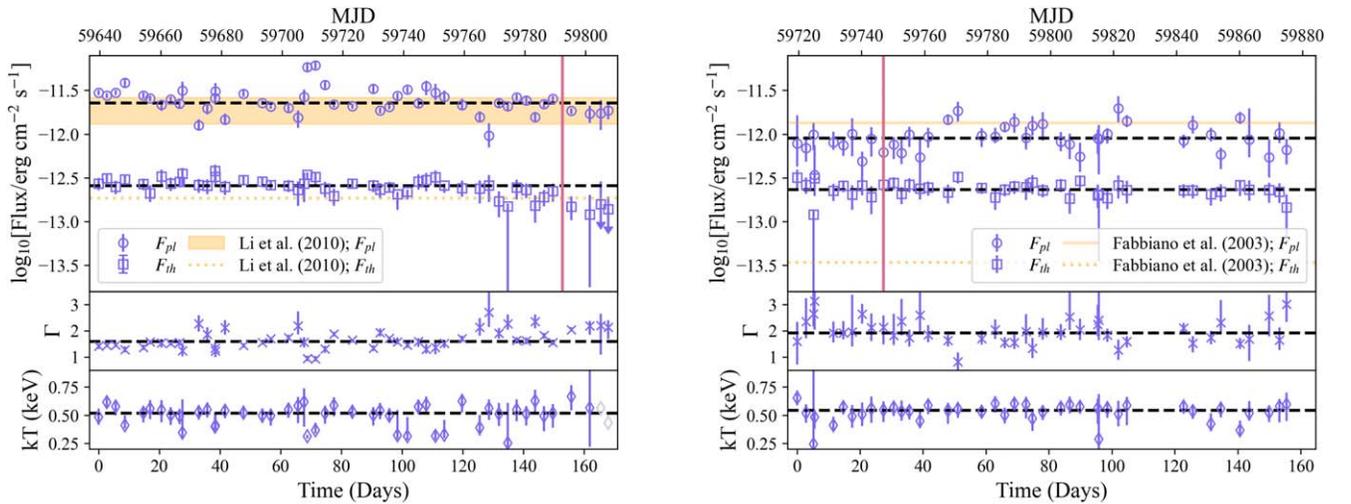

**Figure 2.** Sombrero (left panel) and IC 1459 (right panel) NICER 0.3–8 keV light curves for the $F_{pl}$ (blue circles) and $F_{th}$ (blue squares) components. These are compared with historical Chandra data (orange; power law with solid line, thermal with dashed line) from Z. Li et al. ([2011](#)) and G. Fabbiano et al. ([2003](#)) for Sombrero and IC 1459, respectively. The range of Z. Li et al. ([2011](#))'s Chandra Sombrero $F_{pl}$ data is denoted as a solid orange region. Corresponding $\Gamma$ (blue crosses) and $kT$ (blue diamonds) values are also shown. The last two Sombrero observations have unconstrained $kT$ (gray diamonds) and the $F_{th}$ are therefore marked as upper limits (downward arrows). Median values are shown with black dashed lines. All fitted values are shown with the 90% confidence region. Observation times are reported in modified Julian date (MJD; top x-axis) and zero-centered days (bottom x-axis). The concurrent Swift–NuSTAR observation timing is marked with a vertical solid pink line. Both sources show comparable, roughly constant $F_{th}$ and $kT$, but Sombrero's $F_{pl}$ emission is brighter. Both sources show $F_{pl}$ anticorrelated with $\Gamma$.

In the Swift + NuSTAR broken power-law joint fit of Sombrero, we find a comparable but slightly dimmer $F_{pl}$ than the average found with NICER. The prebreak $\Gamma_1$ is essentially the same as the $\langle \Gamma \rangle$ found with NICER, while the postbreak $\Gamma_2$ is steeper. The fit is insensitive to the thermal component, most likely due to low counts (and by extension, wide binning) in the soft X-ray Swift observation. We therefore freeze $kT$ to the median value from the NICER observations and only estimate an upper limit on $F_{th}$.

Since the thermal component appears roughly constant in the NICER observations, freezing it to the median value should not have a significant impact on the fit results. The upper limit from the joint fit falls close to our median $F_{th}$ found with NICER.

### 3.2. Spectral Modeling: IC 1459

IC 1459's power-law emission is roughly half that of Sombrero, but with comparable thermal emission and $kT$. $\langle \Gamma \rangle$ is slightly steeper than what has been found for Sombrero (but with larger uncertainties). The intrinsic $N_H$ is again $\sim 10^{21}$ atoms cm$^{-2}$, indicating that IC 1459 is also only moderately absorbed.

We compare with a study carried out by G. Fabbiano et al. ([2003](#)): they observe IC 1459 for 60 ks with Chandra and find $F_{pl} \sim 1.34 \times 10^{-12}$ erg cm$^{-2}$ s$^{-1}$, while $F_{th} \sim 0.03 \times 10^{-12}$ erg cm$^{-2}$ s$^{-1}$. The power-law flux is comparable to ours, while the thermal component is slightly lower. As was the case for Sombrero, G. Fabbiano et al. ([2003](#)) find that the combined flux from nonnuclear discrete X-ray sources should be subdominant to $F_{pl}$ and $F_{th}$. For the photon index, G. Fabbiano et al. ([2003](#)) find $\Gamma = 1.88 \pm 0.09$, while O. González-Martín et al. ([2006](#), [2009](#); refitting the existing Chandra data from G. Fabbiano et al. [2003](#) and XMM-Newton data) find $\Gamma = 2.17$ (Chandra) and $\Gamma = 1.89$ (XMM-Newton). This is also in agreement with our $\langle \Gamma \rangle$. G. Fabbiano et al. ([2003](#)) and O. González-Martín et al. ([2006](#), [2009](#)) both report $kT \sim 0.6$ keV and $N_H \sim 10^{21}$ atoms cm$^{-2}$, which agree with our results. L. R. Ivey et al. ([2024](#)) also refit the XMM-Newton data; they freeze the power-law component to the same parameters as G. Fabbiano et al. ([2003](#)) and instead focus on the thermal component, finding $kT \sim 0.7$ keV. This is slightly hotter than our inferred temperature, but they also use a different thermal gas model.

### 3.3. Light Curves: Sombrero

Sombrero's soft X-ray emission seen by NICER (left panel of Figure [2](#)) displays a factor of $\sim 6$ variability in $F_{pl}$, while $F_{th}$ (and the corresponding kT) stays relatively constant. We detect moments of significant variability beyond what has been reported historically, with a brightening period between MJD 59700 and 59720 being particularly notable (see Section 3.6 for further details). The variability in $F_{pl}$ also corresponds to changes in $\Gamma$: we find $\Gamma$ covers a range from ∼0.9 to 2.6.

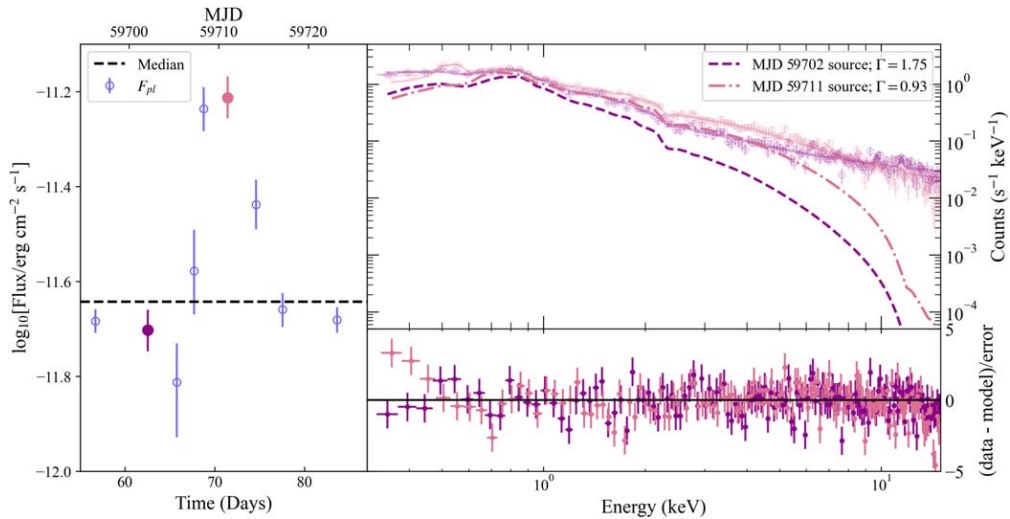

**Figure 3.** Comparison of Sombrero NICER spectra before and during a brightening period between MJD 59700–59720. Left panel: zoomed-in light curve tracking $F_{pl}$ (blue points), with the two highlighted observations shown in purple and pink. The median flux is shown with a dashed black line. Right panel: The corresponding spectra for the highlighted observations. Total counts are shown with circles and the fitted source spectra are shown before (purple dashed line) and during the peak brightening (pink dashed–dotted line). Residuals are shown at the bottom. As $F_{pl}$ increases, the spectral shape stays the same below 2 keV, but the >2 keV slope hardens (i.e., $\Gamma$ decreases).

We compare the observed Sombrero NICER fluxes in Figure 2 with the historical range of Chandra fluxes observed by Z. Li et al. (2011); they find that Sombrero's nuclear power-law emission varies from ∼1.3–2.6 × $10^{-12}$ erg cm$^{-2}$ s$^{-1}$ (a factor of 2).[24] Historical measurements were taken from a handful of X-ray observations spanning a decade, and so it is not surprising that our dedicated monitoring captures more variation in $F_{pl}$ compared to previous works.

### 3.4. Light Curves: IC 1459

IC 1459 shows variability in $F_{pl}$ to a lesser extent than Sombrero, with a roughly constant $F_{th}$ and kT (right panel of Figure 2). $F_{pl}$ varies by a factor of ∼2–3. The average $F_{pl} \approx 10^{-12}$ erg cm$^{-2}$ s$^{-1}$ is consistent with previous Chandra observations by G. Fabbiano et al. (2003). In contrast, G. Fabbiano et al. (2003) find a significantly lower $F_{th}$ (as shown in Figure 2); it is only within the error bars of some of our faintest observations. G. Fabbiano et al. (2003) do not provide uncertainties on their reported fluxes, but the fact that their fitted kT is comparable to ours (∼0.5 keV) alleviates possible model discrepancy concerns.

### 3.5. Light-curve Variability

We address the light-curve variability using the normalized excess variance statistic ($\sigma^2_{NEV}$). It is defined in K. Nandra et al. (1997) as:

$$\sigma^2_{NEV} = \frac{1}{N\mu^2}\sum_{i=1}^{N}[(X_i - \mu)^2 - \sigma_i^2], \quad (3)$$

where $X_i$ is the count rate (or flux) for $N$ points in the light curve, with errors $\sigma_i$, and the unweighted mean count rate (or flux) $\mu$ (see K. Nandra et al. 1997 for further discussion of excess variance and estimating its uncertainty). A slightly altered definition is also provided in F. Vagnetti et al. (2016). We calculate both versions and find them to be equivalent. We find $\sigma^2_{NEV} = 0.12 \pm 0.05$ for Sombrero and $-0.02 \pm 0.05$ (consistent with zero variance beyond the flux uncertainties) for IC 1459. In F. Vagnetti et al. (2016), they estimate $\sigma^2_{NEV}$ for a sample of luminous quasars, the dimmest of which is $L_X \sim 10^{42}$ erg s$^{-1}$ (0.5–4.5 keV). For comparison, our LLAGN targets have $L_X \sim 10^{40}$–$10^{41}$ erg s$^{-1}$ (0.3–8 keV) using the average soft X-ray fluxes measured from NICER.[25] Despite the difference in brightness, Sombrero's $\sigma^2_{NEV}$ is significant and comparable to the values shown in F. Vagnetti et al. (2016). This also agrees with the findings of N. Ding et al. (2018): they identify a handful of cosmologically distant ($z \gtrsim 0.08$) LLAGN candidates with $L_X < 10^{42}$ erg s$^{-1}$ (0.5–7 keV) and find $\sigma^2_{NEV} \sim 0.1-1$.

### 3.6. Interplay between Flux and Spectral Slope

Both Sombrero and IC 1459 show $\Gamma$ varying inversely with $F_{pl}$. The anticorrelation is highlighted in Figure 3, where we zoom in on the MJD 59700–59720 window and compare the spectra before and during its brightening period. Below ∼2 keV, the spectral shape stays the same at both times, consistent with $F_{th}$ and kT remaining roughly constant. As $\Gamma$ increases, however, $F_{pl}$ also increases due to enhanced X-ray emission beyond ∼2 keV. In this energy regime, the emission is dominated by the nuclear power-law component rather than the

---

[24] They simultaneously fit all historical observations with one $\Gamma$, so we cannot compare variability in that parameter.

[25] These are fluxes/luminosities absorbed by the host galaxy, but our fitted $N_H$ values are low enough that this should not significantly impact the inferred luminosities.

**Table 3**
The Reduced $\chi^2$ and Best-fitting Parameters (with $1\sigma$ Errors) for Linear Fits to $\Gamma$ vs. $\log(L_{X,2-10\ keV}/L_{Edd})$ for Sombrero and IC 1459, with M81, NGC 3998, and NGC 3147 Swift Data from S. D. Connolly et al. (2016) Included for Comparison

| Object | $\chi^2$/DOF | Gradient | Intercept | $sp_r$ |
|---|---|---|---|---|
| **Sombrero** | 0.72 | $-1.32 \pm 0.05$ | $-7.08 \pm 0.35$ | $-0.97$ |
| **IC 1459** | 1.12 | $-2.06 \pm 0.14$ | $-11.80 \pm 0.96$ | $-0.94$ |
| M81 | 0.86 | $-0.73 \pm 0.06$ | $-2.12 \pm 0.34$ | $-0.91$ |
| NGC 3998 | 0.95 | $-2.51 \pm 0.39$ | $-12.08 \pm 2.23$ | $-0.78$ |
| NGC 3147 | 0.81 | $-1.21 \pm 0.66$ | $-4.34 \pm 3.16$ | $-0.50$ |
| Combined Avg. | 2.33 | $-0.24 \pm 0.19$ | $0.49 \pm 1.09$ | $-0.50$ |

**Note.** The Spearman's rank correlation coefficient ($sp_r$) for each object is also shown. The best fit to the average fluxes of all included sources is also reported for comparison. The two sources studied in this work are in bold.

constant thermal component. The same inverse trend appears for the total 2–10 keV flux ($F_{X,2-10\ keV}$). We quantify the level of anticorrelation by calculating the Spearman's rank coefficient $sp_r$, provided in Table 3. We find $sp_r = -0.97$ and $sp_r = -0.94$ for Sombrero and IC 1459, respectively, with corresponding p-values of $<10^{-20}$ for both sources, indicating a very strong $F_{X,2-10\ keV}$–$\Gamma$ anticorrelation. The anticorrelation, also known as "harder-when-brighter" behavior, is expected for LLAGNs with certain types of RIAFs, as explained in G. Younes et al. (2011). We discuss this scenario further in Section 4.1.

## 4. Discussion

### 4.1. Flux–Slope Anticorrelation

In the LLAGN RIAF scenario, the Eddington ratio $f_{Edd} \equiv L_{bol}/L_{Edd} \propto L_{X,2-10\ keV}/L_{Edd}$ is inversely proportional to $\Gamma$ (M. Gu & X. Cao 2009). We can convert the NICER flux ($F_{X,2-10\ keV}$) to a luminosity ($L_{X,2-10\ keV}$) and calculate the corresponding $L_{Edd}$ using the measured black hole distances and masses provided in Table 1. In Figure 4, we track $\Gamma$ as a function of $L_{X,2-10\ keV}/L_{Edd}$ for Sombrero and IC 1459 as well as several other individual sources and broad surveys from the literature. Given that $L_{X,2-10\ keV}$ can be rescaled to $L_{bol}$ using a roughly constant factor of $\sim$15–20 for LLAGNs with $L_{X,2-10\ keV} \lesssim 10^{42}$ erg s$^{-1}$ (L. C. Ho 2008; G. Younes et al. 2011; F. Duras et al. 2020), we can treat $L_{X,2-10\ keV}/L_{Edd}$ as a proxy for $f_{Edd}$ without loss of generality. The best-fit line to the data tells us about the gradient of the anticorrelation; we use the orthogonal distance regression method provided by scipy to calculate the best fit taking into account the uncertainties along both axes. Gradient and intercept values and associated uncertainties are provided in Table 3. Both Sombrero and IC 1459 show the expected anticorrelation trend in $L_{X,2-10\ keV}/L_{Edd}$.

This anticorrelation behavior can also be compared with several existing LLAGN studies. We focus our discussion on the very low $f_{Edd}$ ($L_{X,2-10\ keV}/L_{Edd} \lesssim 10^{-5}$) regime, since this is where our targets fall and it is the most undersampled $f_{Edd}$ regime. For examples of "harder-when-brighter" behavior in higher $f_{Edd}$ AGNs, see, e.g., D. Emmanoulopoulos et al. (2012), B. Trakhtenbrot et al. (2017), Y. Diaz et al. (2023), and A. Jana et al. (2023).

First, we consider the Palomar Swift observations collected between 2005 and 2014 for the lowest $f_{Edd}$ sources in S. D. Connolly et al. (2016). We select the subset of these AGNs for which S. D. Connolly et al. (2016) performed linear fits: NGC 3031 (M81), NGC 3998, and NGC 3147.[26] These provide the closest comparison with our targets; M81 and NGC 3998 in particular are among the lowest $f_{Edd}$ and the most well-monitored (in soft X-rays) LLAGNs to date. We convert the reported fluxes to $L_{X,2-10\ keV}/L_{Edd}$ using distances and black hole masses from K. Gültekin et al. (2019) and refit the Swift data using the same fitting routine described above. Our fit results (see Table 3) are comparable to the findings in S. D. Connolly et al. (2016). Our sources show similar anticorrelations to NGC 3998, and slightly steeper trends than for M81. Because our sources have even fainter $L_{X,2-10\ keV}/L_{Edd}$ than those in S. D. Connolly et al. (2016), we can now extend the steep anticorrelation trend seen for individually monitored LLAGNs down to $\sim 10^{-7-8}$.

We compare these well-monitored sources with broader LLAGN surveys. Early works by M. Gu & X. Cao (2009) and A. Constantin et al. (2009) find shallow anticorrelation trends from compiled observations of dozens of LINERs and Seyferts assumed to be hosting LLAGNs. Similar behavior is seen in the LLAGN samples from G. Younes et al. (2011), L. Hernández-García et al. (2013), I. Jang et al. (2014), Q.-X. Yang et al. (2015), and R. She et al. (2018). All the individually monitored sources shown in Figure 4 intersect with the best-fit lines found by these surveys, though the surveys' lines have shallower slopes. The shallow slope could be due to the fact that all the survey data come from snapshot compilations of sources; the best-fit survey lines may be more of an "average" fit to a heterogeneous LLAGN sample, with potential systematic scatter in the data caused by observing with different instruments and at different times. This would also explain why the surveys' observed anticorrelations are often weaker ($sp_r \lesssim -0.7$). A similar situation could explain the minimal slopes found in recent multi-instrument LLAGN sample studies by A. Jana et al. (2023) and Y. Diaz et al. (2023).

To further investigate this hypothesis, we fit another line to the average fluxes from all the monitored sources. The combined average line (shown in red in Figure 4) matches very closely with all the broad surveys (gray lines). Aggregate samples containing snapshots of many sources therefore seem to wash out the steep anticorrelation trend, which only becomes apparent when monitoring individual LLAGNs. Sombrero and IC 1459 are now among the lowest $f_{Edd}$ sources where this anticorrelation is unambiguously detected, and our work verifies the general trends found in other studies of individual sources at higher $f_{Edd}$ (e.g., D. Emmanoulopoulos et al. 2012; S. D. Connolly et al. 2016; S. Niu et al. 2023).

The anticorrelation observed for LLAGNs may be due to an increase in the optical depth of the RIAF with increasing mass accretion rate (and by extension, increasing $L_{X,2-10\ keV}$ and $f_{Edd}$). This leads to an increase in the Compton y-parameter, which traces how many times photons are being scattered and gaining energy. Increasing this parameter results in a harder X-ray spectrum (smaller $\Gamma$) as photons are bumped up to higher energies (M. Gu & X. Cao 2009; L. C. Ho 2009; G. Younes et al. 2011; A. Niedźwiecki et al. 2015).

---

[26] NGC 3147 has only three flux-binned Swift data points, making it difficult to confidently fit a linear model. The fit parameters are therefore more uncertain, and the $sp_r$ is much lower compared to the other monitored sources (see Table 3).

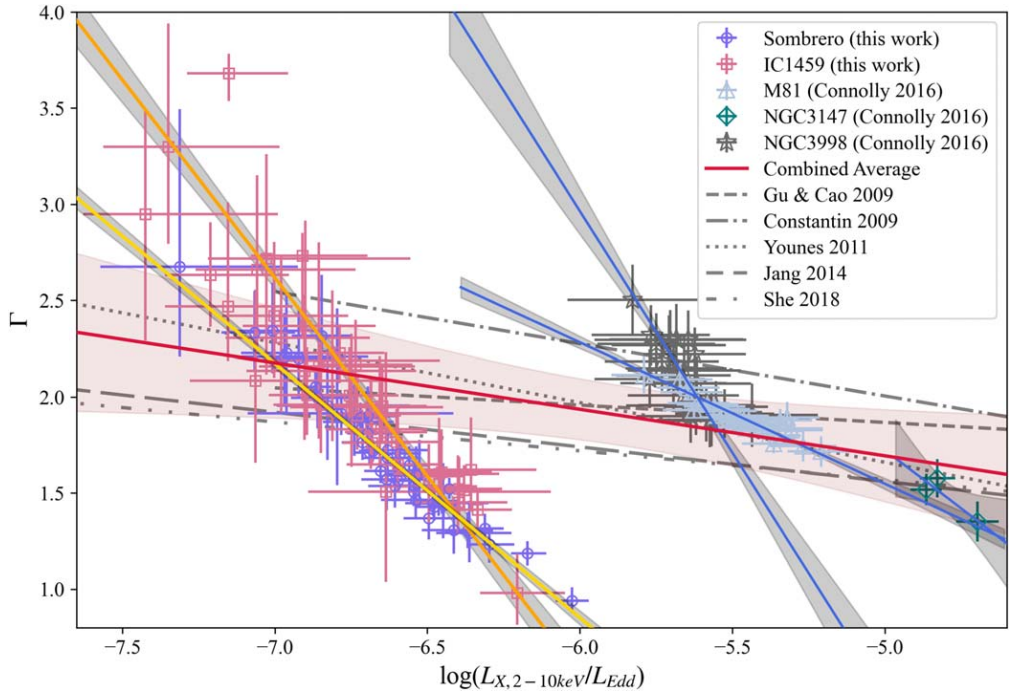

**Figure 4.** Plot of Γ against $\log(L_{X,2-10\,keV}/L_{Edd})$ (as a proxy for $f_{Edd}$) for Sombrero (purple circles), IC 1459 (pink squares), and the lowest $f_{Edd}$ monitored LLAGNs from S. D. Connolly et al. (2016; M81 with blue triangles, NGC 3998 with gray stars, and NGC 3147 with green diamonds). We show our best-fit lines and 1σ confidence intervals (gray shading) for Sombrero (yellow solid line), IC 1459 (orange solid line), and the other monitored sources (blue solid lines); the best-fit parameters are listed in Table 3. We also show the best fits (gray dotted and dashed lines) from several broad LLAGN surveys (A. Constantin et al. 2009; M. Gu & X. Cao 2009; G. Younes et al. 2011; I. Jang et al. 2014; R. She et al. 2018). For comparison with these surveys, we fit a line (red) to the average fluxes of the monitored sources. Sombrero and IC 1459 are now among the SMBHs with the lowest $f_{Edd}$ where the Γ anticorrelation is unambiguously detected.

Interestingly, theoretical works predict there may be a shift from the X-ray emission being RIAF to jet dominated below a threshold of $L_{X,2-10\,keV}/L_{Edd} \sim 10^{-6}$ (F. Yuan & W. Cui 2005; Q. Wu et al. 2007; L. C. Ho 2008; F. Yuan et al. 2009; Q.-X. Yang et al. 2015; S. D. Connolly et al. 2016; A. Jana et al. 2023). Q.-X. Yang et al. (2015) in particular claim that the X-rays at $L_{X,2-10\,keV}/L_{Edd} \lesssim 10^{-6.5}$ come from jet synchrotron emission and the anticorrelation trend should flatten (i.e., Γ should stay constant) in this lowest $f_{Edd}$ regime. We instead find the steep anticorrelation continues below this threshold, extending to at least $L_{X,2-10\,keV}/L_{Edd} \sim 10^{-7}$. While we rule out a jet-induced flattening of the anticorrelation trend, it may still be possible for jets to produce the "harder-when-brighter" behavior if shocks in the jet accelerate electrons and trigger higher-energy SSC scatterings (see S. D. Connolly et al. 2016; J. A. Fernández-Ontiveros et al. 2023 for further discussion). Given that our sources fall very close to the theoretical RIAF/jet-dominated $f_{Edd}$ threshold, our NICER monitoring data could be used in future detailed broadband SED modeling (see, e.g., P. F. Hopkins et al. 2009; B. Bandyopadhyay et al. 2019) to constrain the jet versus RIAF emission contributions.

### 4.2. Power-law Break

Our best-fit power-law break occurs at ∼5–7 keV and is flatter prebreak. This type of break has been observed before in AGN, particularly in certain blazars and quasars (e.g., the NuSTAR + Swift studies in M. Hayashida et al. 2015; V. S. Paliya et al. 2016). These studies both find broken power-law fits with flatter prebreak slopes ($\Gamma_1 \lesssim 1.6, \Gamma_2 \gtrsim 1.7$) and $E_{break} \sim 2-10$ keV. They attribute the break to inverse Compton emission via SSC and/or external Compton.

Interestingly, our $E_{break}$ is also similar to the results found from fitting NuSTAR observations of M87* during the 2018 EHT campaign (J. C. Algaba et al. 2024). In their case $E_{break} \sim 10$ keV, but $\Gamma_1$ is steeper than $\Gamma_2$, whereas we find the reverse. M87* is an LLAGN like our sources; its properties are summarized in Table 1. In J. C. Algaba et al. (2024), they test multiple broadband SED models for M87* with and without an X-ray break. The break appears in their model at the transition region between jet-powered synchrotron and SSC dominating the emission. Sombrero also has jets that dominate the radio/submillimeter emission (see summary in Section 1), but it is unclear how much they contribute to X-ray emission (see Z. Li et al. 2011; J. A. Fernández-Ontiveros et al. 2023). Detailed broadband SED modeling for Sombrero using our X-ray data would be useful to constrain the origins of the fitted break, as would deeper observations in the hard X-ray regime.

### 4.3. Sombrero Variability Timescale

We perform a statistical analysis of Sombrero's light curve to search for periodicity and the characteristic variability timescale. Due to slight variations in the NICER monitoring cadence and removal of poor quality observations, the time between observations ultimately ranges from 0.13 to 6.9 days with an average of 2.9 days. We must therefore employ timing analysis method(s) that can handle data which are unevenly sampled in time. The Lomb–Scargle periodogram is a least-squares method for analyzing unevenly sampled astronomical time series data. It works by iteratively fitting a sinusoidal model to the light curve at a range of frequencies, with a larger power reflecting a better fit at that frequency (J. T. VanderPlas 2018). We use the Lomb–Scargle tool provided by `astropy`, but find that the light curve is white-noise-dominated and shows no strong evidence for significant periodicity.

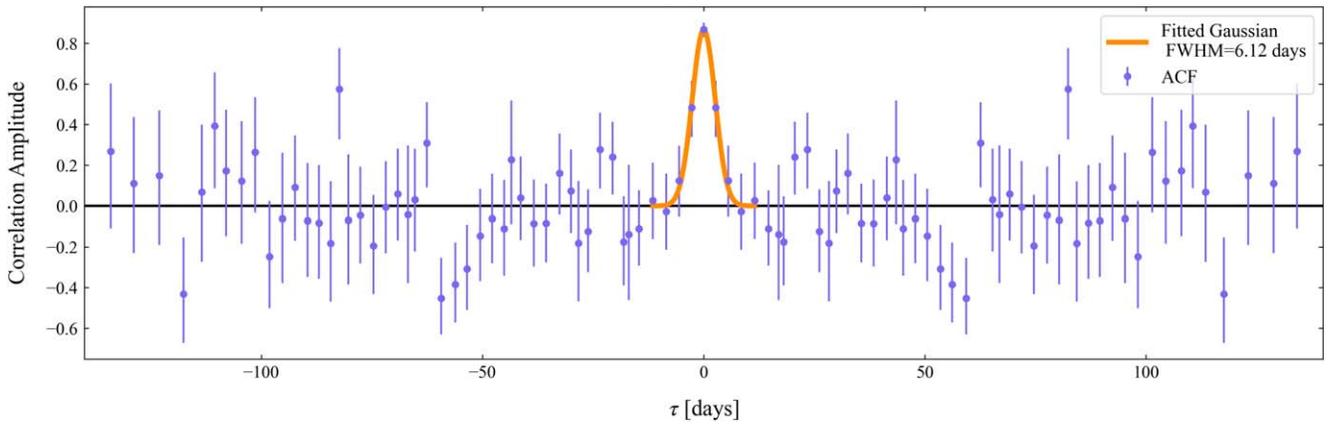

**Figure 5.** ACF results from the Sombrero NICER light curve shown in Figure 2. We consider timescales ranging from 0 to 150 days. The set of nonzero correlation amplitudes (blue points) centered around timescale $\tau = 0$ are treated as the first peak in the ACF. A Gaussian fit to this peak (orange line) has an FWHM $\tau = 6.12$ days. This peak width can be interpreted as the shortest variability timescale in the light curve.

The characteristic timescale for aperiodic variability can also be estimated using the correlation function. We use the `pyzdcf` package.[27] It is designed for robust estimation of the cross-correlation—or autocorrelation—function of sparse and unevenly sampled astronomical time series. We applied an autocorrelation function (ACF) to our Sombrero NICER light curve, with the results shown in Figure 5. Inspired by the methods of T. Hovatta et al. (2007), we identify the peak centered at zero lag. We fit a Gaussian to this peak and estimate its FWHM: for Sombrero, we find a timescale $\tau = 6.12$ days. Taking into account the uncertainty on the correlation function amplitudes, which was estimated using 100 Monte Carlo simulations, $\tau$ is somewhere between 5 and 7 days. This is close to the predicted Nyquist frequency for our average cadence ($2 \times 2.9$ days $\sim 5.8$ days), and so any variability timescale below this limit is not well resolved. As seen in Figures 2 and 3, there is a factor of 4 variability in $F_{pl}$ on a timescale faster than 3.5 days. So, significant variability is happening on short timescales $\lesssim 6$ days, but we can still treat our inferred autocorrelation timescale as an upper limit.

Sombrero's aperiodic variability timescale is comparable to theoretical predictions for material orbiting at $<100\ r_g$ from the central SMBH. Two of the shortest—and by extension closest to the SMBH—theoretical timescales that we can probe observationally are the light-travel timescale and the freefall timescale. The light-crossing time is given by:

$$\tau_{lc} = \frac{r}{c}. \quad (4)$$

Meanwhile, the radial freefall timescale for the simplified case of spherical adiabatic Bondi accretion is:

$$\tau_{ff} \approx \left(\frac{r^3}{GM}\right)^{1/2}. \quad (5)$$

For Sombrero's central SMBH (see Table 1 for source properties) at the radius of the innermost stable circular orbit (ISCO; $r_{ISCO}$; 3 $r_g$ for a nonspinning BH), we find $\tau_{lc} \sim 0.2$ days and $\tau_{ff} \sim 0.6$ days. If we equate the 6 day autocorrelation timescale with $\tau_{lc}$ or $\tau_{ff}$ and solve for $r$, this corresponds to $r_{lc} \sim 78\ r_g$ and $r_{ff} \sim 15\ r_g$. We are therefore not able to probe the ISCO with our data, but we do see variability on rapid timescales corresponding to an emission region with $r < 100\ r_g$. In a study by X. Yan et al. (2024), they report an observed 88 GHz nuclear core size of $\sim 25 \pm 5\ r_g$—on the same order of magnitude as our inferred $r_{lc}$ and $r_{ff}$—that is consistent with their model predictions from an advection dominated accretion flow (ADAF). They also fit Sombrero's broadband SED with a coupled ADAF–jet model and find that the ADAF dominates at both submillimeter and X-ray frequencies.

## 5. Conclusions

In this study, we perform X-ray monitoring of two LLAGNs, Sombrero and IC 1459. We use NICER to track variations in the soft X-ray emission, and demonstrate NICER's joint source–background modeling capabilities to analyze very faint AGNs in combination with NuSTAR and Swift. From these observations we find the following.

1. Both sources are generally described by a combination of power-law and thermal components. The average spectral fit parameters agree with literature values for both targets, with Sombrero's power-law emission being slightly brighter than IC 1459's. Both sources have a thermal component with a temperature of $\sim 0.5$ keV, with IC 1459 having a slightly softer power-law photon index $\Gamma$ than Sombrero ($\langle \Gamma \rangle = 2.1$ and 1.7, respectively).
2. For Sombrero, there is tentative evidence for a spectral break in the NuSTAR data at $\sim 6$ keV. This is similar to findings for several other AGNs, where it could correspond to synchrotron/inverse Compton emission in the soft X-ray band.
3. Sombrero shows a factor of $\sim 6$ flux variability in the power-law emission, significantly more than what has been seen historically. The variability shows a characteristic timescale of $\lesssim 5$–7 days. The variability timescale indicates the nuclear X-ray emission likely originates at $<100\ r_g$.
4. Both Sombrero and IC 1459 are "harder when brighter." They are among the lowest $f_{Edd}$ sources to exhibit this behavior. This steep anticorrelation trend appears in several other well-monitored LLAGNs. If all of these sources' $\langle\Gamma\rangle$ and $\langle L_X/L_{Edd}\rangle$ are fit together, the anticorrelation trend becomes much shallower, as seen in many previous LLAGN surveys.

---
[27] `pyzdcf` is a Python module that emulates a widely used Fortran program called the z-transformed discrete correlation function (ZDCF; T. Alexander 1997).

Coordinated multiwavelength monitoring of LLAGNs is essential to search for correlated variability and constrain the emission region properties. As we demonstrate in this work, Sombrero is of particular interest due to its significant variability and soft X-ray brightness sufficient for monitoring with NICER. IC 1459 also shows interesting $\Gamma$ anticorrelation behavior, and may be a more suitable target for other X-ray instruments such as XMM-Newton and Chandra. Future EHT observations of both these LLAGNs provide an urgent motivation for coordinated monitoring, since both submillimeter and X-ray bands are predicted to be either RIAF dominated or have comparable jet–RIAF contributions (X. Yan et al. 2024). Meanwhile, upcoming deeper NuSTAR + NICER observations (PI: N. M. Ford, PID 10244) will help constrain the spectral break(s) while revealing any changes in the hard X-ray emission. Given the level of variability seen in Sombrero with NICER, quasi-simultaneous multiwavelength observations will be essential to inform future time-resolved broadband SED studies aimed at disentangling jet versus RIAF emission contributions. Expanding high-cadence monitoring to a sample of LLAGNs with instruments such as NICER will be a crucial next step in characterizing both the variability and the SED shape, and determining how different AGN physical properties (e.g., black hole mass, Eddington ratio, and absorption/obscuration) impact the observed emission.

## Acknowledgments

N.M.F. is from O'ahu, one of the islands of Hawai'i, an indigenous space where the descendants of the original people are kānaka 'ōiwi/Native Hawaiian. N.M.F. is a visitor in Tiohtiá:ke/Mooniyang, also called Montréal, which lies on the unceded lands of the Kanien'kehá:ka. We are grateful to the traditional stewards of the lands, waters, and skies that we gather within. As we engage in astronomy on these lands, we respect, listen to, and make space for indigenous voices and ways of knowing the sky and Earth that sustain us all.

This research has made use of the NASA/IPAC Extragalactic Database (NED), which is funded by the National Aeronautics and Space Administration and operated by the California Institute of Technology.

The authors gratefully acknowledge Neil Nagar for his work in defining a sample of high priority EHT targets. We also thank Sera Markoff and Ruby Duncan for their insights into the theoretical modeling of LLAGN broadband SEDs and emission mechanisms. The authors appreciate Craig Markwardt for his development of the SCORPEON background model and for his guidance on applying SCORPEON to our data.

N.M.F. and D.H. acknowledge funding from the Natural Sciences and Engineering Research Council of Canada (NSERC) Discovery Grant program and the Canada Research Chairs (CRC) program. N.M.F. acknowledges funding from the Fondes de Recherche Nature et Technologies (FRQNT) Doctoral research scholarship. N.M.F. and D.H. acknowledge support from the Canadian New Frontiers in Research Fund (NFRF)—Explorations program and the Trottier Space Institute at McGill. The authors acknowledge support from the Centre de recherche en astrophysique du Québec, un regroupement stratégique du FRQNT. V.R. acknowledges the financial support from the visitor and mobility program of the Finnish Centre for Astronomy with ESO (FINCA), funded by the Academy of Finland grant number 306531. D.G.N. acknowledges funding from ANID through Fondecyt Postdoctorado (project code 3220195) and Nucleo Milenio TITANs (project NCN2022002).

*Facilities:* NICER, NuSTAR, Swift (XRT).

*Software:* astropy (Astropy Collaboration et al. 2013, 2018, 2022), scipy (P. Virtanen et al. 2020), HEASoft, XSPEC (K. A. Arnaud 1996).

## Appendix
## NICER Spectral Fitting Results

Tables of NICER observation spectral fitting results for Sombrero (Table 4) and IC 1459 (Table 5) discussed in the main text.

**Table 4**
Fitted Spectral Properties of the NICER Observations of Sombrero

| ObsID | $F_{\rm pl}^{\rm a}$ | $F_{\rm th}^{\rm a}$ | $F_{\rm X,2-10\,keV}$ | $\Gamma$ | $kT$ | $N_{\rm H}^{\rm b}$ | C-stat. | DOF | Reduced |
| | log(Flux/erg cm$^{-2}$ s$^{-1}$) | | | | (keV) | ($10^{22}$ atoms cm$^{-2}$) | | | C-stat. |
|---|---|---|---|---|---|---|---|---|---|
| 5676020101 | $-11.53_{-0.03}^{+0.03}$ | $-12.57_{-0.07}^{+0.08}$ | $-11.59_{-0.04}^{+0.04}$ | $1.42_{-0.08}^{+0.11}$ | $0.48_{-0.08}^{+0.07}$ | ... | 175 | 132 | 1.33 |
| 5676020201 | $-11.56_{-0.03}^{+0.06}$ | $-12.51_{-0.06}^{+0.05}$ | $-11.63_{-0.04}^{+0.05}$ | $1.44_{-0.07}^{+0.07}$ | $0.61_{-0.04}^{+0.05}$ | ... | 151 | 131 | 1.16 |
| 5676020301 | $-11.53_{-0.03}^{+0.03}$ | $-12.60_{-0.07}^{+0.08}$ | $-11.60_{-0.05}^{+0.05}$ | $1.46_{-0.08}^{+0.08}$ | $0.58_{-0.05}^{+0.08}$ | ... | 163 | 136 | 1.20 |
| 5676020401 | $-11.41_{-0.04}^{+0.05}$ | $-12.52_{-0.06}^{+0.08}$ | $-11.46_{-0.05}^{+0.06}$ | $1.28_{-0.13}^{+0.11}$ | $0.41_{-0.09}^{+0.03}$ | ... | 181 | 138 | 1.31 |
| 5676020601 | $-11.56_{-0.04}^{+0.04}$ | $-12.57_{-0.06}^{+0.07}$ | $-11.62_{-0.05}^{+0.05}$ | $1.37_{-0.10}^{+0.10}$ | $0.52_{-0.06}^{+0.08}$ | ... | 142 | 139 | 1.02 |
| 5676020701 | $-11.59_{-0.03}^{+0.03}$ | $-12.67_{-0.09}^{+0.10}$ | $-11.71_{-0.06}^{+0.04}$ | $1.58_{-0.07}^{+0.10}$ | $0.56_{-0.08}^{+0.09}$ | ... | 162 | 124 | 1.31 |
| 5676020801 | $-11.67_{-0.06}^{+0.07}$ | $-12.48_{-0.08}^{+0.12}$ | $-11.78_{-0.11}^{+0.09}$ | $1.54_{-0.15}^{+0.18}$ | $0.55_{-0.09}^{+0.11}$ | ... | 106 | 107 | 1.00 |
| 5676020901 | $-11.60_{-0.03}^{+0.04}$ | $-12.57_{-0.07}^{+0.09}$ | $-11.69_{-0.05}^{+0.05}$ | $1.52_{-0.10}^{+0.10}$ | $0.50_{-0.07}^{+0.09}$ | ... | 140 | 121 | 1.16 |
| 5676021001 | $-11.65_{-0.06}^{+0.06}$ | $-12.54_{-0.06}^{+0.07}$ | $-11.75_{-0.09}^{+0.09}$ | $1.55_{-0.12}^{+0.12}$ | $0.49_{-0.06}^{+0.07}$ | ... | 148 | 133 | 1.12 |
| 5676021002 | $-11.50_{-0.10}^{+0.10}$ | $-12.45_{-0.06}^{+0.14}$ | $-11.60_{-0.13}^{+0.14}$ | $1.24_{-0.28}^{+0.22}$ | $0.34_{-0.30}^{+0.05}$ | ... | 124 | 108 | 1.15 |
| 5676021201 | $-11.90_{-0.07}^{+0.06}$ | $-12.58_{-0.07}^{+0.09}$ | $-12.21_{-0.11}^{+0.12}$ | $2.27_{-0.32}^{+0.30}$ | $0.53_{-0.06}^{+0.07}$ | $0.09_{-0.05}^{+0.04}$ | 258 | 142 | 1.82 |
| 5676021301 | $-11.71_{-0.08}^{+0.07}$ | $-12.59_{-0.07}^{+0.10}$ | $-11.89_{-0.11}^{+0.11}$ | $1.87_{-0.27}^{+0.25}$ | $0.55_{-0.07}^{+0.07}$ | $0.05_{-0.04}^{+0.04}$ | 128 | 130 | 0.99 |
| 5676021401 | $-11.59_{-0.06}^{+0.06}$ | $-12.42_{-0.07}^{+0.13}$ | $-11.64_{-0.08}^{+0.09}$ | $1.33_{-0.20}^{+0.21}$ | $0.40_{-0.13}^{+0.05}$ | ... | 117 | 111 | 1.06 |
| 5676021402 | $-11.51_{-0.09}^{+0.08}$ | $-12.49_{-0.08}^{+0.11}$ | $-11.56_{-0.10}^{+0.12}$ | $1.23_{-0.21}^{+0.24}$ | $0.41_{-0.10}^{+0.04}$ | ... | 135 | 128 | 1.06 |
| 5676021501 | $-11.83_{-0.07}^{+0.06}$ | $-12.60_{-0.07}^{+0.09}$ | $-12.01_{-0.10}^{+0.09}$ | $2.13_{-0.29}^{+0.27}$ | $0.54_{-0.05}^{+0.07}$ | $0.07_{-0.04}^{+0.04}$ | 154 | 131 | 1.18 |
| 5676021701 | $-11.54_{-0.04}^{+0.03}$ | $-12.53_{-0.05}^{+0.06}$ | $-11.61_{-0.06}^{+0.05}$ | $1.45_{-0.08}^{+0.10}$ | $0.52_{-0.06}^{+0.05}$ | ... | 173 | 141 | 1.23 |
| 5676021901 | $-11.64_{-0.03}^{+0.03}$ | $-12.54_{-0.05}^{+0.06}$ | $-11.74_{-0.04}^{+0.04}$ | $1.55_{-0.08}^{+0.08}$ | $0.50_{-0.06}^{+0.06}$ | ... | 133 | 128 | 1.04 |
| 5676022001 | $-11.68_{-0.03}^{+0.02}$ | $-12.58_{-0.06}^{+0.06}$ | $-11.83_{-0.04}^{+0.04}$ | $1.70_{-0.06}^{+0.08}$ | $0.50_{-0.06}^{+0.09}$ | ... | 254 | 128 | 1.99 |
| 5676022201 | $-11.70_{-0.04}^{+0.04}$ | $-12.59_{-0.07}^{+0.07}$ | $-11.86_{-0.07}^{+0.07}$ | $1.75_{-0.10}^{+0.11}$ | $0.55_{-0.06}^{+0.07}$ | ... | 131 | 133 | 0.99 |
| 5676022301 | $-11.81_{-0.08}^{+0.12}$ | $-12.64_{-0.10}^{+0.18}$ | $-12.11_{-0.17}^{+0.18}$ | $2.19_{-0.56}^{+0.33}$ | $0.58_{-0.08}^{+0.16}$ | $0.08_{-0.08}^{+0.04}$ | 152 | 138 | 1.11 |
| 5676022401 | $-11.58_{-0.09}^{+0.09}$ | $-12.57_{-0.13}^{+0.21}$ | $-11.67_{-0.12}^{+0.13}$ | $1.57_{-0.19}^{+0.21}$ | $0.62_{-0.12}^{+0.22}$ | ... | 150 | 114 | 1.32 |
| 5676022402 | $-11.24_{-0.05}^{+0.05}$ | $-12.46_{-0.02}^{+0.05}$ | $-11.32_{-0.05}^{+0.05}$ | $0.94_{-0.11}^{+0.11}$ | $0.31_{-0.01}^{+0.02}$ | ... | 243 | 148 | 1.65 |
| 5676022501 | $-11.21_{-0.05}^{+0.04}$ | $-12.49_{-0.05}^{+0.09}$ | $-11.17_{-0.05}^{+0.05}$ | $0.93_{-0.11}^{+0.10}$ | $0.37_{-0.07}^{+0.03}$ | ... | 178 | 133 | 1.34 |
| 5676022601 | $-11.44_{-0.05}^{+0.05}$ | $-12.62_{-0.08}^{+0.10}$ | $-11.44_{-0.07}^{+0.08}$ | $1.30_{-0.13}^{+0.14}$ | $0.52_{-0.07}^{+0.10}$ | ... | 219 | 139 | 1.58 |
| 5676022701 | $-11.66_{-0.03}^{+0.04}$ | $-12.71_{-0.08}^{+0.11}$ | $-11.87_{-0.07}^{+0.06}$ | $1.87_{-0.09}^{+0.09}$ | $0.59_{-0.07}^{+0.10}$ | ... | 146 | 129 | 1.13 |
| 5676022901 | $-11.68_{-0.03}^{+0.03}$ | $-12.57_{-0.05}^{+0.05}$ | $-11.80_{-0.05}^{+0.05}$ | $1.64_{-0.05}^{+0.09}$ | $0.53_{-0.06}^{+0.06}$ | ... | 191 | 127 | 1.51 |
| 5676023101 | $-11.48_{-0.05}^{+0.05}$ | $-12.59_{-0.06}^{+0.08}$ | $-11.51_{-0.07}^{+0.07}$ | $1.33_{-0.12}^{+0.11}$ | $0.50_{-0.07}^{+0.09}$ | ... | 149 | 131 | 1.14 |
| 5676023201 | $-11.73_{-0.03}^{+0.03}$ | $-12.64_{-0.07}^{+0.10}$ | $-11.93_{-0.06}^{+0.06}$ | $1.93_{-0.16}^{+0.14}$ | $0.54_{-0.07}^{+0.08}$ | $0.05_{-0.03}^{+0.02}$ | 167 | 129 | 1.30 |
| 5676023301 | $-11.69_{-0.03}^{+0.03}$ | $-12.61_{-0.06}^{+0.06}$ | $-11.84_{-0.04}^{+0.04}$ | $1.71_{-0.07}^{+0.08}$ | $0.49_{-0.07}^{+0.06}$ | ... | 216 | 127 | 1.70 |
| 5676023401 | $-11.56_{-0.03}^{+0.06}$ | $-12.69_{-0.02}^{+0.18}$ | $-11.74_{-0.06}^{+0.06}$ | $1.58_{-0.17}^{+0.10}$ | $0.32_{-0.22}^{+0.04}$ | ... | 144 | 130 | 1.11 |
| 5676023501 | $-11.49_{-0.04}^{+0.04}$ | $-12.66_{-0.02}^{+0.09}$ | $-11.58_{-0.03}^{+0.04}$ | $1.46_{-0.12}^{+0.11}$ | $0.31_{-0.17}^{+0.04}$ | ... | 149 | 126 | 1.19 |
| 5676023601 | $-11.65_{-0.07}^{+0.05}$ | $-12.54_{-0.08}^{+0.09}$ | $-11.69_{-0.07}^{+0.13}$ | $1.59_{-0.13}^{+0.20}$ | $0.57_{-0.06}^{+0.06}$ | ... | 149 | 124 | 1.20 |
| 5676023701 | $-11.46_{-0.07}^{+0.10}$ | $-12.53_{-0.08}^{+0.13}$ | $-11.51_{-0.12}^{+0.12}$ | $1.30_{-0.20}^{+0.16}$ | $0.59_{-0.07}^{+0.08}$ | ... | 153 | 127 | 1.21 |
| 5676023801 | $-11.53_{-0.10}^{+0.08}$ | $-12.49_{-0.08}^{+0.09}$ | $-11.68_{-0.11}^{+0.12}$ | $1.37_{-0.27}^{+0.27}$ | $0.32_{-0.05}^{+0.03}$ | ... | 140 | 133 | 1.06 |
| 5676023901 | $-11.57_{-0.07}^{+0.07}$ | $-12.59_{-0.02}^{+0.09}$ | $-11.78_{-0.12}^{+0.08}$ | $1.53_{-0.07}^{+0.18}$ | $0.33_{-0.13}^{+0.03}$ | ... | 166 | 140 | 1.19 |
| 5676024101 | $-11.67_{-0.07}^{+0.06}$ | $-12.62_{-0.09}^{+0.10}$ | $-11.78_{-0.09}^{+0.09}$ | $1.70_{-0.12}^{+0.14}$ | $0.62_{-0.06}^{+0.10}$ | ... | 153 | 136 | 1.13 |
| 5676024301 | $-11.81_{-0.07}^{+0.06}$ | $-12.62_{-0.09}^{+0.17}$ | $-12.07_{-0.11}^{+0.13}$ | $2.13_{-0.36}^{+0.32}$ | $0.39_{-0.11}^{+0.07}$ | $0.10_{-0.06}^{+0.04}$ | 143 | 143 | 1.00 |
| 5676024401 | $-12.02_{-0.14}^{+0.13}$ | $-12.59_{-0.12}^{+0.19}$ | $-12.46_{-0.28}^{+0.39}$ | $2.70_{-0.90}^{+0.55}$ | $0.56_{-0.08}^{+0.10}$ | $0.11_{-0.13}^{+0.08}$ | 163 | 136 | 1.21 |
| 5676024501 | $-11.64_{-0.05}^{+0.05}$ | $-12.77_{-0.11}^{+0.18}$ | $-11.84_{-0.08}^{+0.09}$ | $1.90_{-0.21}^{+0.19}$ | $0.51_{-0.10}^{+0.11}$ | $0.04_{-0.03}^{+0.03}$ | 133 | 126 | 1.06 |
| 5676024601 | $-11.68_{-0.04}^{+0.05}$ | $-12.83_{-0.03}^{+1.09}$ | $-11.99_{-0.09}^{+0.11}$ | $2.27_{-0.35}^{+0.23}$ | $0.25_{-0.36}^{+0.11}$ | $0.10_{-0.05}^{+0.04}$ | 149 | 129 | 1.16 |
| 5676024701 | $-11.58_{-0.04}^{+0.04}$ | $-12.61_{-0.09}^{+0.14}$ | $-11.71_{-0.06}^{+0.06}$ | $1.64_{-0.11}^{+0.12}$ | $0.55_{-0.09}^{+0.11}$ | ... | 135 | 117 | 1.16 |
| 5676024801 | $-11.62_{-0.05}^{+0.05}$ | $-12.64_{-0.07}^{+0.09}$ | $-11.69_{-0.06}^{+0.07}$ | $1.62_{-0.12}^{+0.13}$ | $0.51_{-0.07}^{+0.10}$ | ... | 187 | 130 | 1.44 |
| 5676024901 | $-11.81_{-0.06}^{+0.06}$ | $-12.82_{-0.12}^{+0.20}$ | $-12.15_{-0.13}^{+0.14}$ | $2.35_{-0.28}^{+0.25}$ | $0.63_{-0.10}^{+0.10}$ | $0.06_{-0.04}^{+0.03}$ | 182 | 140 | 1.30 |
| 5676025001 | $-11.66_{-0.04}^{+0.04}$ | $-12.72_{-0.08}^{+0.11}$ | $-11.84_{-0.08}^{+0.07}$ | $1.82_{-0.09}^{+0.12}$ | $0.49_{-0.09}^{+0.14}$ | ... | 137 | 122 | 1.13 |
| 5676025101 | $-11.59_{-0.03}^{+0.03}$ | $-12.65_{-0.09}^{+0.12}$ | $-11.69_{-0.05}^{+0.05}$ | $1.56_{-0.10}^{+0.11}$ | $0.52_{-0.08}^{+0.13}$ | ... | 126 | 111 | 1.14 |
| 5676025301 | $-11.73_{-0.04}^{+0.04}$ | $-12.83_{-0.12}^{+0.15}$ | $-12.01_{-0.05}^{+0.05}$ | $2.05_{-0.05}^{+0.07}$ | $0.67_{-0.08}^{+0.12}$ | ... | 192 | 113 | 1.70 |
| 5676025501 | $-11.77_{-0.11}^{+0.11}$ | $-12.92_{-0.23}^{+0.83}$ | $-12.12_{-0.21}^{+0.17}$ | $2.20_{-0.21}^{+0.34}$ | $0.57_{-0.34}^{+0.35}$ | ... | 98 | 99 | 0.99 |
| 5676025602 | $-11.76_{-0.15}^{+0.19}$ | $<-12.53$ | $-11.94_{-0.27}^{+0.48}$ | $2.20_{-0.45}^{+1.10}$ | ... | ... | 117 | 88 | 1.34 |
| 5676025701 | $-11.73_{-0.09}^{+0.10}$ | $<-12.72$ | $-11.98_{-0.21}^{+0.21}$ | $2.13_{-0.28}^{+0.44}$ | ... | ... | 98 | 90 | 1.09 |

**Notes.**
[a] $F_{\rm pl}$ and $F_{\rm th}$ are for the 0.3–8 keV energy range.
[b] The majority of the fits are insensitive to the intrinsic $N_{\rm H}$; in these cases, we freeze $N_{\rm H}$ to the median value from all fits (see Table 2).

Table 5
Fitted Spectral Properties of the NICER Observations of IC 1459

| ObsID | $F_{\rm pl}^{a}$ | $F_{\rm th}^{a}$ | $F_{\rm X, 2-10\ keV}$ | $\Gamma$ | $kT$ | $N_{\rm H}^{b}$ | C-stat. | DOF | Reduced C-stat. |
|---|---|---|---|---|---|---|---|---|---|
| | log(Flux/erg cm$^{-2}$ s$^{-1}$) | | | | (keV) | ($10^{22}$ atoms cm$^{-2}$) | | | |
| 5676010101 | $-12.11^{+0.27}_{-0.32}$ | $-12.50^{+0.14}_{-0.09}$ | $-12.14^{+0.38}_{-0.35}$ | $1.59^{+0.87}_{-0.75}$ | $0.65^{+0.06}_{-0.07}$ | ... | 128 | 115 | 1.12 |
| 5676010201 | $-12.16^{+0.15}_{-0.13}$ | $-12.58^{+0.10}_{-0.08}$ | $-12.51^{+0.21}_{-0.18}$ | $2.35^{+0.63}_{-0.89}$ | $0.52^{+0.09}_{-0.07}$ | $0.12^{+0.09}_{-0.15}$ | 139 | 121 | 1.16 |
| 5676010301 | $-12.00^{+0.18}_{-0.13}$ | $-12.92^{+1.76}_{-0.04}$ | $-12.42^{+0.21}_{-0.16}$ | $2.65^{+0.48}_{-0.82}$ | $0.24^{+0.11}_{-0.87}$ | $0.16^{+0.08}_{-0.23}$ | 117 | 103 | 1.14 |
| 5676010302 | $-12.47^{+0.24}_{-0.34}$ | $-12.50^{+0.16}_{-0.06}$ | $-12.66^{+0.16}_{-0.15}$ | $3.14^{+1.10}_{-2.86}$ | $0.48^{+0.10}_{-0.06}$ | $1.04^{+0.57}_{-2.01}$ | 135 | 116 | 1.17 |
| 5676010501 | $-12.09^{+0.09}_{-0.12}$ | $-12.65^{+0.12}_{-0.08}$ | $-12.24^{+0.11}_{-0.10}$ | $1.91^{+0.36}_{-0.44}$ | $0.41^{+0.05}_{-0.08}$ | $0.08^{+0.06}_{-0.08}$ | 152 | 126 | 1.21 |
| 5676010601 | $-12.13^{+0.12}_{-0.06}$ | $-12.59^{+0.07}_{-0.06}$ | $-12.40^{+0.14}_{-0.13}$ | $1.92^{+0.25}_{-0.45}$ | $0.57^{+0.07}_{-0.04}$ | $0.04^{+0.04}_{-0.08}$ | 159 | 123 | 1.30 |
| 5676010701 | $-12.00^{+0.26}_{-0.18}$ | $-12.69^{+0.17}_{-0.11}$ | $-12.32^{+0.28}_{-0.25}$ | $1.96^{+0.56}_{-1.42}$ | $0.49^{+0.10}_{-0.10}$ | $0.10^{+0.09}_{-0.38}$ | 119 | 123 | 0.97 |
| 5676010801 | $-12.31^{+0.24}_{-0.11}$ | $-12.59^{+0.11}_{-0.08}$ | $-12.72^{+0.18}_{-0.19}$ | $2.63^{+0.37}_{-0.39}$ | $0.51^{+0.10}_{-0.11}$ | ... | 191 | 137 | 1.40 |
| 5676010901 | $-12.05^{+0.20}_{-0.13}$ | $-12.72^{+0.19}_{-0.12}$ | $-12.36^{+0.25}_{-0.23}$ | $2.11^{+0.51}_{-0.55}$ | $0.56^{+0.13}_{-0.11}$ | $0.07^{+0.06}_{-0.16}$ | 117 | 134 | 0.88 |
| 5676011001 | $-12.21^{+0.29}_{-0.22}$ | $-12.58^{+0.12}_{-0.09}$ | $-12.41^{+0.24}_{-0.26}$ | $2.13^{+0.64}_{-0.45}$ | $0.55^{+0.11}_{-0.07}$ | ... | 143 | 135 | 1.06 |
| 5676011101 | $-12.12^{+0.17}_{-0.09}$ | $-12.56^{+0.08}_{-0.06}$ | $-12.36^{+0.19}_{-0.15}$ | $1.83^{+0.37}_{-0.82}$ | $0.57^{+0.07}_{-0.05}$ | $0.08^{+0.07}_{-0.18}$ | 135 | 125 | 1.08 |
| 5676011201 | $-12.22^{+0.12}_{-0.16}$ | $-12.68^{+0.12}_{-0.07}$ | $-12.49^{+0.23}_{-0.21}$ | $2.37^{+0.55}_{-0.85}$ | $0.53^{+0.08}_{-0.10}$ | $0.15^{+0.09}_{-0.13}$ | 137 | 130 | 1.06 |
| 5676011301 | $-12.01^{+0.14}_{-0.09}$ | $-12.58^{+0.09}_{-0.07}$ | $-12.26^{+0.17}_{-0.12}$ | $1.75^{+0.34}_{-0.29}$ | $0.54^{+0.08}_{-0.05}$ | ... | 180 | 133 | 1.36 |
| 5676011401 | $-12.27^{+0.28}_{-0.22}$ | $-12.62^{+0.13}_{-0.09}$ | $-12.19^{+0.10}_{-0.09}$ | $2.60^{+0.73}_{-2.13}$ | $0.45^{+0.06}_{-0.09}$ | $0.30^{+0.27}_{-1.08}$ | 134 | 114 | 1.18 |
| 5676011501 | $-12.03^{+0.13}_{-0.10}$ | $-12.61^{+0.09}_{-0.07}$ | $-12.19^{+0.17}_{-0.14}$ | $1.86^{+0.29}_{-0.36}$ | $0.59^{+0.08}_{-0.05}$ | $0.09^{+0.05}_{-0.09}$ | 120 | 115 | 1.05 |
| 5676011701 | $-11.83^{+0.05}_{-0.06}$ | $-12.67^{+0.12}_{-0.09}$ | $-11.97^{+0.18}_{-0.15}$ | $1.64^{+0.24}_{-0.21}$ | $0.54^{+0.12}_{-0.07}$ | ... | 127 | 119 | 1.07 |
| 5676011801 | $-11.73^{+0.12}_{-0.10}$ | $-12.49^{+0.08}_{-0.06}$ | $-11.71^{+0.14}_{-0.12}$ | $0.83^{+0.33}_{-0.35}$ | $0.56^{+0.08}_{-0.05}$ | ... | 114 | 118 | 0.97 |
| 5676012001 | $-12.02^{+0.13}_{-0.09}$ | $-12.61^{+0.07}_{-0.05}$ | $-12.10^{+0.21}_{-0.09}$ | $1.70^{+0.23}_{-0.32}$ | $0.53^{+0.05}_{-0.06}$ | $0.09^{+0.05}_{-0.08}$ | 125 | 126 | 0.99 |
| 5676012201 | $-12.03^{+0.07}_{-0.10}$ | $-12.73^{+0.14}_{-0.10}$ | $-12.15^{+0.35}_{-0.10}$ | $2.05^{+0.35}_{-0.39}$ | $0.60^{+0.08}_{-0.07}$ | $0.12^{+0.07}_{-0.07}$ | 142 | 128 | 1.11 |
| 5676012301 | $-11.92^{+0.06}_{-0.05}$ | $-12.63^{+0.08}_{-0.06}$ | $-12.00^{+0.09}_{-0.08}$ | $1.56^{+0.22}_{-0.19}$ | $0.50^{+0.08}_{-0.06}$ | ... | 137 | 120 | 1.14 |
| 5676012401 | $-11.86^{+0.13}_{-0.10}$ | $-12.60^{+0.09}_{-0.07}$ | $-11.96^{+0.17}_{-0.12}$ | $1.56^{+0.24}_{-0.32}$ | $0.60^{+0.08}_{-0.05}$ | $0.06^{+0.05}_{-0.09}$ | 140 | 120 | 1.17 |
| 5676012501 | $-12.04^{+0.13}_{-0.13}$ | $-12.61^{+0.14}_{-0.09}$ | $-12.16^{+0.17}_{-0.15}$ | $1.97^{+0.45}_{-0.67}$ | $0.60^{+0.13}_{-0.06}$ | $0.18^{+0.11}_{-0.16}$ | 130 | 123 | 1.06 |
| 5676012601 | $-11.90^{+0.11}_{-0.11}$ | $-12.55^{+0.13}_{-0.09}$ | $-11.91^{+0.16}_{-0.12}$ | $1.34^{+0.38}_{-0.39}$ | $0.47^{+0.11}_{-0.09}$ | ... | 124 | 117 | 1.06 |
| 5676012701 | $-11.89^{+0.19}_{-0.14}$ | $-12.64^{+0.08}_{-0.07}$ | $-12.24^{+0.13}_{-0.14}$ | $1.94^{+0.27}_{-0.55}$ | $0.53^{+0.07}_{-0.06}$ | $0.06^{+0.05}_{-0.10}$ | 136 | 124 | 1.10 |
| 5676012901 | $-12.09^{+0.11}_{-0.11}$ | $-12.58^{+0.08}_{-0.06}$ | $-12.29^{+0.10}_{-0.11}$ | $1.91^{+0.27}_{-0.34}$ | $0.57^{+0.07}_{-0.05}$ | ... | 163 | 135 | 1.21 |
| 5676013001 | $-12.12^{+0.17}_{-0.12}$ | $-12.74^{+0.17}_{-0.10}$ | $-12.56^{+0.27}_{-0.24}$ | $2.53^{+0.50}_{-0.96}$ | $0.59^{+0.08}_{-0.07}$ | $0.17^{+0.07}_{-0.16}$ | 124 | 136 | 0.91 |
| 5676013101 | $-12.26^{+0.18}_{-0.15}$ | $-12.54^{+0.08}_{-0.06}$ | $-12.57^{+0.29}_{-0.29}$ | $2.05^{+0.40}_{-0.41}$ | $0.58^{+0.06}_{-0.05}$ | ... | 114 | 132 | 0.87 |
| 5676013203 | $-12.05^{+0.18}_{-0.10}$ | $-12.70^{+0.15}_{-0.12}$ | $-12.41^{+0.27}_{-0.18}$ | $2.24^{+0.43}_{-0.75}$ | $0.56^{+0.14}_{-0.14}$ | $0.12^{+0.07}_{-0.14}$ | 129 | 125 | 1.04 |
| 5676013301 | $-12.06^{+0.40}_{-0.16}$ | $-12.69^{+0.77}_{-0.01}$ | $-12.53^{+0.25}_{-0.32}$ | $2.40^{+0.76}_{-0.32}$ | $0.29^{+0.09}_{-0.40}$ | $0.20^{+0.15}_{-1.26}$ | 130 | 107 | 1.22 |
| 5676013401 | $-12.00^{+0.11}_{-0.06}$ | $-12.73^{+0.12}_{-0.10}$ | $-12.14^{+0.15}_{-0.10}$ | $1.84^{+0.25}_{-0.24}$ | $0.56^{+0.08}_{-0.24}$ | ... | 126 | 113 | 1.12 |
| 5676013501 | $-11.71^{+0.16}_{-0.14}$ | $-12.58^{+0.15}_{-0.11}$ | $-11.86^{+0.18}_{-0.16}$ | $1.26^{+0.37}_{-0.39}$ | $0.51^{+0.11}_{-0.12}$ | ... | 152 | 130 | 1.17 |
| 5676013601 | $-11.85^{+0.07}_{-0.07}$ | $-12.64^{+0.15}_{-0.11}$ | $-11.90^{+0.11}_{-0.10}$ | $1.59^{+0.23}_{-0.21}$ | $0.59^{+0.18}_{-0.08}$ | ... | 131 | 112 | 1.18 |
| 5676014201 | $-12.07^{+0.09}_{-0.07}$ | $-12.64^{+0.09}_{-0.07}$ | $-12.19^{+0.12}_{-0.07}$ | $2.11^{+0.23}_{-0.19}$ | $0.58^{+0.06}_{-0.05}$ | ... | 110 | 121 | 0.91 |
| 5676014301 | $-11.90^{+0.10}_{-0.11}$ | $-12.64^{+0.09}_{-0.07}$ | $-11.84^{+0.08}_{-0.07}$ | $1.53^{+0.35}_{-0.27}$ | $0.54^{+0.07}_{-0.05}$ | ... | 135 | 124 | 1.09 |
| 5676014501 | $-12.01^{+0.08}_{-0.08}$ | $-12.69^{+0.13}_{-0.06}$ | $-12.11^{+0.11}_{-0.10}$ | $1.74^{+0.26}_{-0.25}$ | $0.42^{+0.07}_{-0.09}$ | ... | 137 | 131 | 1.05 |
| 5676014601 | $-12.23^{+0.17}_{-0.08}$ | $-12.66^{+0.11}_{-0.08}$ | $-12.66^{+0.27}_{-0.16}$ | $2.31^{+0.34}_{-0.86}$ | $0.56^{+0.08}_{-0.06}$ | $0.07^{+0.06}_{-0.14}$ | 155 | 123 | 1.27 |
| 5676014801 | $-11.82^{+0.07}_{-0.07}$ | $-12.58^{+0.12}_{-0.06}$ | $-11.89^{+0.08}_{-0.07}$ | $1.51^{+0.21}_{-0.21}$ | $0.36^{+0.04}_{-0.09}$ | ... | 128 | 118 | 1.09 |
| 5676014901 | $-12.06^{+0.20}_{-0.36}$ | $-12.64^{+0.12}_{-0.11}$ | $-12.25^{+0.27}_{-0.22}$ | $1.69^{+0.85}_{-0.54}$ | $0.52^{+0.07}_{-0.07}$ | ... | 145 | 131 | 1.11 |
| 5676015101 | $-12.27^{+0.23}_{-0.20}$ | $-12.64^{+0.15}_{-0.08}$ | $-12.93^{+0.27}_{-0.30}$ | $2.56^{+0.77}_{-0.81}$ | $0.52^{+0.08}_{-0.06}$ | $0.12^{+0.08}_{-0.18}$ | 146 | 132 | 1.11 |
| 5676015201 | $-11.99^{+0.11}_{-0.13}$ | $-12.66^{+0.13}_{-0.10}$ | $-12.00^{+0.14}_{-0.12}$ | $1.65^{+0.37}_{-0.49}$ | $0.58^{+0.14}_{-0.07}$ | $0.08^{+0.07}_{-0.12}$ | 126 | 125 | 1.01 |
| 5676015301 | $-12.18^{+0.17}_{-0.12}$ | $-12.84^{+0.32}_{-0.15}$ | $-12.86^{+0.29}_{-0.26}$ | $3.01^{+0.67}_{-1.07}$ | $0.60^{+0.15}_{-0.10}$ | $0.19^{+0.08}_{-0.16}$ | 121 | 132 | 0.92 |

**Note.**
[a] $F_{\rm pl}$ and $F_{\rm th}$ are for the 0.3–8 keV energy range.
[b] The majority of the fits are insensitive to the intrinsic $N_{\rm H}$; in these cases, we freeze $N_{\rm H}$ to the median value from all fits (see Table 2).

### ORCID iDs

Nicole M. Ford https://orcid.org/0000-0001-8921-3624
Michael Nowak https://orcid.org/0000-0001-6923-1315
Venkatessh Ramakrishnan https://orcid.org/0000-0002-9248-086X
Daryl Haggard https://orcid.org/0000-0001-6803-2138
Kristen Dage https://orcid.org/0000-0002-8532-4025
Dhanya G. Nair https://orcid.org/0000-0001-5357-7805
Chi-kwan Chan https://orcid.org/0000-0001-6337-6126